\documentclass{aa}
\usepackage{txfonts}
\usepackage{graphicx}
\begin{document}

\title{The long-period Galactic Cepheid RS\,Puppis\thanks{Based on observations made with ESO telescopes at La Silla Paranal Observatory, under ESO program 383.D-0623(A).}}
\subtitle{II. 3D structure and mass of the nebula from VLT/FORS polarimetry}
\titlerunning{3D structure and mass of the circumstellar nebula of RS\,Pup}
\authorrunning{Kervella et al.}
\author{
P.~Kervella\inst{1}
\and
A.~M\'erand\inst{2}
\and
L.~Szabados\inst{3}
\and
W.~B.~Sparks\inst{4}
\and
R.~J.~Havlen\inst{5}
\and
H.~E.~Bond\inst{4}
\and
E.~Pompei\inst{2}
\and
P.~Fouqu\'e\inst{6}
\and
D.~Bersier\inst{7}
\and
M.~Cracraft\inst{4}
}
\offprints{Pierre Kervella}
\mail{pierre.kervella@obspm.fr}
\institute{
%LESIA-Observatoire de Paris, CNRS, UPMC Univ. Paris 6, Univ. Paris-Diderot,
LESIA, Observatoire de Paris, CNRS UMR 8109, UPMC, Universit\'e Paris Diderot,
5 place Jules Janssen, F-92195 Meudon, France
\and
European Southern Observatory, Alonso de Cordova 3107,
Casilla 19001, Vitacura, Santiago 19, Chile
\and
Konkoly Observatory, H-1525 Budapest XII, P. O. Box 67, Hungary
\and
Space Telescope Science Institute, 3700 San Martin Drive, Baltimore, MD 21218, USA
\and
307 Big Horn Ridge Dr., Albuquerque, NM 87122, USA
\and
IRAP, UMR 5277, CNRS, Universit\'e de Toulouse, 14 av. E. Belin, F-31400 Toulouse, France
\and
Astrophysics Research Institute, Liverpool John Moores University, Twelve Quays House, Egerton Wharf, Birkenhead, CH411LD, United Kingdom
}
\date{Received ; Accepted}
\abstract
% Context
{The southern long-period Cepheid RS\,Pup is surrounded by a large circumstellar dusty nebula reflecting the light from the central star. Due to the changing luminosity of the central source, light echoes propagate into the nebula. This remarkable phenomenon was the subject of Paper I. The origin and physical properties of the nebula are however uncertain: it may have been created through mass loss from the star itself, or it could be the remnant of a pre-existing interstellar cloud.}
% Aims
{Our goal is to determine the three-dimensional structure of the light-scattering nebula, and estimate its mass. This will bring us new clues on the origin of the nebula. Knowing the geometrical shape of the nebula will also allow us to retrieve the distance of RS\,Pup in an unambiguous manner using a model of its light echoes (in a forthcoming work).}
% Methods
{The scattering angle of the Cepheid light in the circumstellar nebula can be recovered from its degree of linear polarization. We thus observed the nebula surrounding RS\,Pup using the polarimetric imaging mode of the VLT/FORS instrument, and obtained a map of the degree and position angle of linear polarization.}
% Results
{From our FORS observations, we derive a three-dimensional map of the distribution of the dust around RS\,Pup, whose overall geometry is an irregular and geometrically thin layer. The nebula does not present a well-defined central symmetry. Using a simple scattering model, we derive a total dust mass of $M_\mathrm{dust} = 2.9 \pm 0.9\,M_\odot$ for the light-scattering dust within $1.8\arcmin$ of the Cepheid. This translates into a total mass of $M_\mathrm{gas+dust} = 290 \pm 120\,M_\odot$, assuming a dust-to-gas ratio of $M_\mathrm{dust} / M_\mathrm{gas} = 1.0 \pm 0.3\%$.}
% Conclusions
{The high mass of the dusty nebula excludes that it was created by mass-loss from the star. However, the thinness of the dust distribution is an indication that the Cepheid participated to the shaping of the nebula, e.g. through its radiation pressure or stellar wind. RS\,Pup therefore appears as a regular long-period Cepheid located in an exceptionally dense interstellar environment.}

\maketitle

%__________________________________Introduction
\section{Introduction}

RS\,Pup (\object{HD 68860}, \object{SAO 198944}) is the only Cepheid known to be intimately associated with a large circumstellar nebula. From NTT imaging of the light echoes present in the nebula, Kervella et al.~(\cite{kervella08}, hereafter Paper I) derived its distance with a precision of $\pm 1.4$\%. The method employed in that article relies on the assumption that the observed nebular features are located close to the plane of the sky. As argued by Bond \& Sparks~(\cite{bond09}), this assumption, if not correct, could lead to a significant bias on the determined distance. In order to overcome this limitation, the goal of the present work is to determine the three-dimensional geometry of RS\,Pup's nebula. For this purpose, we obtained imaging polarimetry data using the VLT/FORS2 instrument. These data give us accurate maps of the polarization degree and angle over the nebula (Sect.~\ref{fors2obs}). The polarimetric properties of the dust are related to the scattering angle of the light, and thus the position of the scattering dust relative to the plane of the sky. We complete these data by the NTT/EMMI observations presented in Paper~I (Sect.~\ref{emmiobs}). We then estimate the thickness of the light-scattering dust and its three-dimensional distribution (Sect.~\ref{3Dshape}). We discuss the total mass and origin of the nebula in Sect.~\ref{discussion}. In Appendix~\ref{tempevol}, we examine the possibility that the nebular features evolved over the last five decades.

%__________________________________Observations
\section{VLT/FORS2 polarimetric observations\label{fors2obs}}

\subsection{Instrumental setup and observation log \label{log}}

\begin{figure}[]
\centering
\includegraphics[width=\hsize]{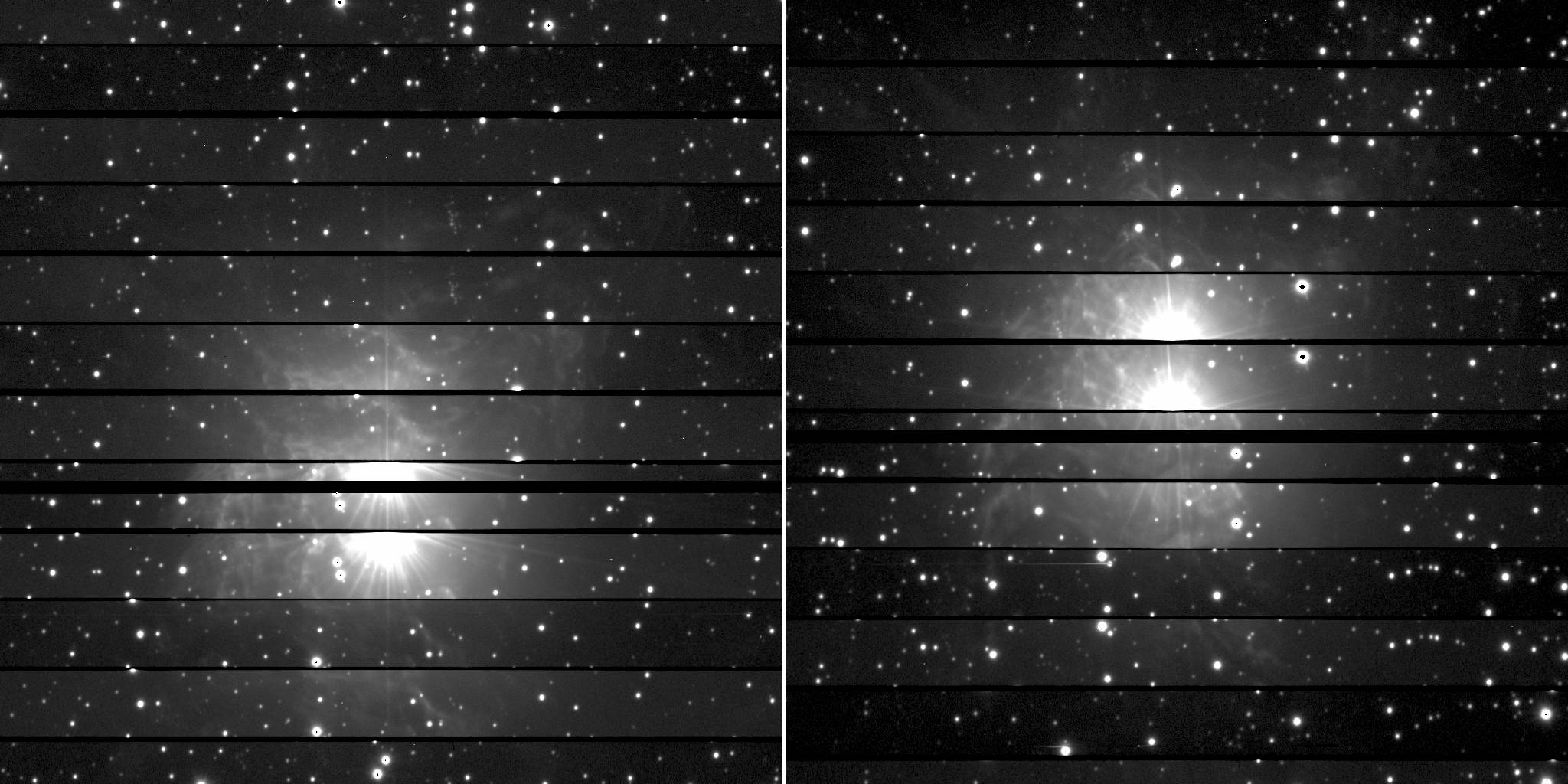}
\caption{Examples of FORS polarimetric exposures of RS\,Pup, each corresponding to one of the four 60\,s exposures labeled \#12 and \#16 in Table~\ref{fors_log}. Each horizontal stripe corresponds alternatively to the ordinary and extraordinary beams. The CCD gap is visible below the center of the field. The field of view is $4.3 \arcmin \times 4.3\arcmin$. \label{stripesExample}}
\end{figure}

We observed RS\,Pup on the nights of 4 April, 26 April and 26 May 2009 using the FOcal Reducer and low dispersion Spectrograph FORS\footnote{http://www.eso.org/sci/facilities/paranal/instruments/fors/} (Appenzeller et al.~\cite{appenzeller98}; O'Brien~\cite{obrien09}). This instrument is the result of the merging of the original FORS1 and FORS2 instruments, and is presently installed at the Cassegrain focus of the Unit Telescope 1 (UT1) of the Very Large Telescope (VLT). The imaging polarimetry with this instrument is obtained by inserting in the light path a half-wave superachromatic retarder plate (hereafter HWP) and a Wollaston prism.
%
%The HWP introduces a 180$^\circ$ phase shift in one linear polarization of the incoming light beam, while leaving the other polarization direction unchanged. The effect of the HWP on the incoming beam is equivalent (for polarimetry) to rotating the whole instrument.
%
The HWP introduces a 180$^\circ$ phase shift on the component of the incoming polarized light which is along its reference axis, leaving the normal component unchanged (e.g. Clarke \& Grainger~\cite{clarke71}). This effectively rotates the incoming polarization position angle (PA) by an amount twice as large as the angle between the incoming PA and the reference axis PA. This is equivalent to rotating the whole instrument, and hence the Wollaston, by a corresponding amount.
The HWP is mounted on a rotating support. The Wollaston prism acts as a beam-splitting analyser, i.e. it splits the light in two orthogonally polarized beams propagating in two directions, separated by an angle of $22\arcsec$ on the sky.
%
% A strip mask is inserted in the beam after the Wollaston to avoid that the two beams overlap on the detector.
A strip mask is inserted in the entrance focal plane and then re-imaged on the image plane detector, to avoid that the two beams overlap.
A single exposure therefore gives a 50\% field coverage only, but with the two polarization directions imaged simultaneously on the detector. This optical setup including a Wollaston has the advantage of being insensitive to changes in the atmospheric turbulence or sky transparency, as the two orthogonally polarized images are obtained simultaneously (Fig.~\ref{stripesExample}).
We covered the RS\,Pup nebula with two sets of exposures (for each epoch), shifted on the sky by $16\arcsec$ to cover most of the nebula. This strategy leaves a series of $6\arcsec$ stripes without observations, giving a total field coverage of approximately 80\%. The reason for leaving this gap around the central star is to avoid its direct imaging on the detector, which would have caused a heavy saturation. To cover the rest of the nebula, the 26 May observation was obtained with the instrument rotated by $90^\circ$. The observations obtained on 26 April are incomplete, with only 3 positions of the HWP. This leaves two complete epochs, 4 April and 26 May, that we will refer to as ``first" and ``second" epochs in the following.

%______________ Figure

We obtained four 60\,seconds exposures for each position of the HWP (0$^\circ$, 22.5$^\circ$, 45$^\circ$, 67.5$^\circ$), resulting in a total exposure time of 960\,s per telescope offset position, or 1920\,s per complete epoch. The complete list of exposures is given in Table~\ref{fors_log}. The phases $\phi$ were computed using the reference epoch and period listed in Paper~I: $T_0 = \mathrm{MJD}\,54090.836$ and $P = 41.4389\,\mathrm{days}$. We used the $V_{\rm high}$ filter $\#114$ of FORS, that is a broader version of the standard Bessell $V$ band filter (effective wavelength of 557\,nm, FWHM of 123.5\,nm). The pixel scale was $0.252\arcsec$/binned pixel, and the detector was read in the standard mode (2x2 pixels binning). We completed our data set with the observation of a PSF calibrator, Alphard (\object{HD 81797}, \object{HR 3748}), in order to measure the scattered light halo in the instrument and subtract it from the images of RS\,Pup (Sect.~\ref{datared}). The exposure time was set to 0.6\,s to match the ADU counts level of RS\,Pup on the detector. Alphard was imaged at the same position as RS\,Pup on the detector, to obtain the same illumination of the array.

% RS Pup position: 08:13 = 123.25¡, -34:34=-34.56¡
% Moon position 2009-04-05: 139.008¡, 15.696¡, separation=53¡
% Moon position 2009-04-26: 57.531¡, 25.269¡, separation=86¡
% Moon position 2009-05-26: 100.207¡, 25.497¡, separation=64¡

%__________________________________Table of observations
\begin{table}
\caption{Log of the FORS observations of RS\,Pup.}
\label{fors_log}
\begin{tabular}{llccllr}
\hline \hline
%\begin{small}
$\#$ & UT date (2009) & MJD & $\phi$ & $\rho(\arcsec)$ & AM & $\alpha$($^\circ$) \\
\hline
\noalign{\smallskip}
& {\it RS Pup} \\
1 & 04/04 23:54:55 & 54925.997 & 0.1540 & 1.33 & 1.02 & 0.0 \\
2 & 05/04 00:00:49 & 54926.001 & 0.1541 & 1.23 & 1.02 & 22.5 \\
3 & 05/04 00:06:44 & 54926.005 & 0.1542 & 1.06 & 1.02 & 45.0 \\
4 & 05/04 00:12:39 & 54926.008 & 0.1543 & 1.05 & 1.02 & 67.5 \\
5 & 05/04 00:18:54 & 54926.013 & 0.1544 & 1.05 & 1.02 & 0.0 \\
6 & 05/04 00:24:49 & 54926.017 & 0.1545 & 0.97 & 1.02 & 22.5 \\
7 & 05/04 00:30:44 & 54926.021 & 0.1546 & 1.15 & 1.02 & 45.0 \\
8 & 05/04 00:36:39 & 54926.026 & 0.1547 & 1.07 & 1.03 & 67.5 \\
\noalign{\smallskip}
9 & 26/04 23:49:08 & 54947.992 & 0.6848 & 1.28 & 1.06 & 0.0 \\
10 & 26/04 23:55:41 & 54947.997 & 0.6849 & 1.45 & 1.07 & 22.5 \\
11 & 27/04 00:00:33 & 54948.000 & 0.6850 & 1.89 & 1.07 & 45.0 \\
\noalign{\smallskip}
12 & 26/05 23:16:53 & 54977.970 & 0.4082 & 0.98 & 1.24 & 0.0 \\
13 & 26/05 23:22:46 & 54977.974 & 0.4083 & 0.73 & 1.26 & 22.5 \\
14 & 26/05 23:28:39 & 54977.978 & 0.4084 & 0.69 & 1.28 & 45.0 \\
15 & 26/05 23:34:32 & 54977.982 & 0.4085 & 0.80 & 1.30 & 67.5 \\
16 & 26/05 23:40:48 & 54977.987 & 0.4086 & 0.90 & 1.32 & 0.0 \\
17 & 26/05 23:46:42 & 54977.991 & 0.4087 & 0.73 & 1.35 & 22.5 \\
18 & 26/05 23:52:36 & 54977.995 & 0.4088 & 0.70 & 1.38 & 45.0 \\
19 & 26/05 23:58:29 & 54977.999 & 0.4089 & 0.73 & 1.40 & 67.5 \\
\hline
\noalign{\smallskip}
& {\it Alphard} \\
20 & 26/05 23:59:12 & 54977.999 &  & 0.72 & 1.41 & 67.5 \\
21 & 27/05 00:13:59 & 54978.010 &  & 0.87 & 1.26 & 0.0 \\
22 & 27/05 00:15:29 & 54978.011 &  & 0.83 & 1.27 & 22.5 \\
23 & 27/05 00:16:59 & 54978.012 &  & 0.77 & 1.27 & 45.0 \\
24 & 27/05 00:18:29 & 54978.013 &  & 0.79 & 1.28 & 67.5 \\
25 & 27/05 00:20:21 & 54978.014 &  & 0.75 & 1.29 & 0.0 \\
26 & 27/05 00:21:51 & 54978.015 &  & 0.78 & 1.29 & 22.5 \\
27 & 27/05 00:23:21 & 54978.016 &  & 0.83 & 1.30 & 45.0 \\
28 & 27/05 00:24:51 & 54978.017 &  & 0.73 & 1.30 & 67.5 \\
\hline
%\end{small}
\end{tabular}
\tablefoot{MJD is the average modified julian date of the exposures, and $\alpha$ is the position angle of the HWP. The seeing $\rho$ in the visible measured by the observatory DIMM is also listed, as well as the observation airmass (AM).}
\end{table}

\subsection{Data preprocessing \label{datared}}

The individual raw images were pre-processed  in a standard way (bias subtraction, flat-fielding, bad pixel masking) using the IRAF\footnote{IRAF is distributed by the NOAO, which are operated by the Association of Universities for Research in Astronomy, Inc., under cooperative agreement with the National Science Foundation.} software package. The flat-field was obtained without the Wollaston prism. As discussed by Patat \& Romaniello~(\cite{patat06}), the flat-fielding uncertainties are mitigated by the introduction of some redundancy in the polarimetric measurements (we used four HWP positions). We checked that the rotation of the HWP does not induce any measurable displacement of the image on the detector. A significant differential geometric distorsion however exists between the stripes corresponding to the ordinary and extraordinary beams. The differential image displacement reaches approximately $0.25\arcsec$ (1 pixel) horizontally, and $1.25\arcsec$ (5 pixels) vertically. To compensate for this image distorsion, we mapped the ordinary and extraordinary beam images to the same fiducial image from our 2008 observations with the NTT/EMMI instrument (Paper~I). As astrometric references, we selected a dozen stars over the image, and we applied a polynomial geometrical transformation using the IRAF {\tt wregister} command. The resulting RMS dispersion of the star positions relative to the transformed world coordinate system (WCS) is approximately $0.12\arcsec$, or half a camera pixel, which is acceptable.

The observations of Alphard were pre-processed in the same way as RS\,Pup's. Thanks to the very short exposures, very few background stars are present in the images. An exponential model of the instrumental/atmospheric scattered light halo was derived from the Alphard images using ring median estimates of the radial dependence of the halo intensity. This modeling is necessary to avoid the contamination of the RS Pup images by the diffraction spikes present in the Alphard images at a different position angle. We used the halo model to generate synthetic halo images for the ordinary and extraordinary beams. They were then scaled to the brightness of RS\,Pup at the epoch of each observation, and subtracted from the added beam images (ordinary plus extraordinary stripes) before the computation of the Stokes parameters (Sect.~\ref{stokesparam}). 

\subsection{Polarimetric quantities \label{stokesparam}}

Following Goldman et al.~(\cite{goldman09}), we obtained the degree and position angle of linear polarization by calculating the normalized Stokes parameters $p_Q = Q/I$ and $p_U = U/I$ (see also Patat \& Romaniello~\cite{patat06}, and Zapatero Osorio et al.~\cite{zapatero05}). They were derived from the processed FORS images using the following expressions:
\begin{equation}
F(\alpha)= \frac{f_{\rm O}(\alpha) - f_{\rm E}(\alpha)}{{\rm O}(\alpha)+f_{\rm E}(\alpha)} \\
\end{equation}
\begin{equation}
p_Q = Q/I = 1/2\ \left[F(0^\circ) - F(45^\circ)\right] \\
\end{equation}
\begin{equation}
p_U = U/I = 1/2\ \left[F(22.5^\circ) - F(67.5^\circ)\right] \\
\end{equation}
\begin{equation}
p_{\rm L} = \sqrt{p_Q^2 + p_U^2}
\end{equation}
where $p_{\rm L}$ is the degree of linear polarization, and $f_{\rm O}(\alpha)$ and $f_{\rm E}(\alpha)$ are respectively the ordinary and extraordinary fluxes obtained at the retarder angle $\alpha$. In this expression, $p_{\rm L}$ carries a bias due to the fact that it is a positive definite quantity. It was therefore debiased using Table~A1 from Sparks \& Axon~(\cite{sparks99}).
A detailed discussion on this particular question was also presented by Simmons \& Stewart~(\cite{simmons85}). The statistical uncertainties of the measurements were estimated from the measured standard deviation $\sigma(F)$ of the $F(\alpha)$ values derived from the four exposures obtained at each position angle of the HWP. The propagation of the $F(\alpha)$ errors for the degree of linear polarization $p_{\rm L}$ gives the variance $\sigma^2(p_{\rm L})$:
\begin{equation}
\sigma^2(p_{\rm _L}) = \frac{p_Q^2\,\sigma^2(p_Q) + p_U^2\,\sigma^2(p_U)}{p_Q^2 + p_U^2}
\end{equation}
The values of $\sigma^2$ for each pixel in the $p_Q$ and $p_U$ images were derived from the variance of each pixel values in our sequence of four $F$ images obtained at the same HWP position angle. As the accuracy of this method is limited by the small number of $F$ exposures, there is a relatively large statistical dispersion on the $\sigma^2$ values. We therefore chose to average the $\sigma^2(p_{\rm _L})$ for each pixel over its surrounding $3 \times 3$\,pixel box to derive more accurate estimates of the actual variance. We checked that the actual variance of the $p_{\rm L}$ pixel values over the 9 pixel box is in agreement with this computed average value, giving us confidence in the resulting error bars.
The polarization angle $\theta$ and its associated variance $\sigma^2(\theta)$ (for statistically significant polarization degrees) are derived using the following expressions (Landi Degl'Innocenti et al.~\cite{landi07}):
\begin{equation}
\theta = \frac{1}{2}\ {\rm sign}(p_U)\ \arccos \left( \frac{p_Q}{\sqrt{p_Q^2 + p_U^2}} \right),
\end{equation}
\begin{equation}
\sigma^2(\theta) = \frac{\sigma^2(p_U)}{|p_Q + p_U^2/p_Q|} + \frac{p_U^2\ \sigma^2(p_Q)}{ (p_Q^2 + p_U^2)^2}.
\end{equation}
The uncertainties we obtain are $\sigma(p_{\rm L}) \approx 1$ to 5\% and $\sigma(\theta) \approx 0.03$ to 0.10\,radians (2 to $6^\circ$) within $1\arcmin$ of RS\,Pup (sections of the nebula with an average surface brightness).

As pointed out by Patat \& Romaniello~(\cite{patat06}), FORS is affected by instrumental polarization close to the edge of the detector. However, the nebula surrounding RS\,Pup is contained within the central $3\times3$\,arcminutes (including its faint extensions), where the instrumental polarization is limited to approximately 0.5\%. Considering the small amplitude of this bias, we neglect it in our analysis.

\subsection{Combination of epochs 1 and 2}

%______________ Figure
\begin{figure*}[ht]
\centering
% 952 pixel images = 4.0' field
\includegraphics[width=9cm]{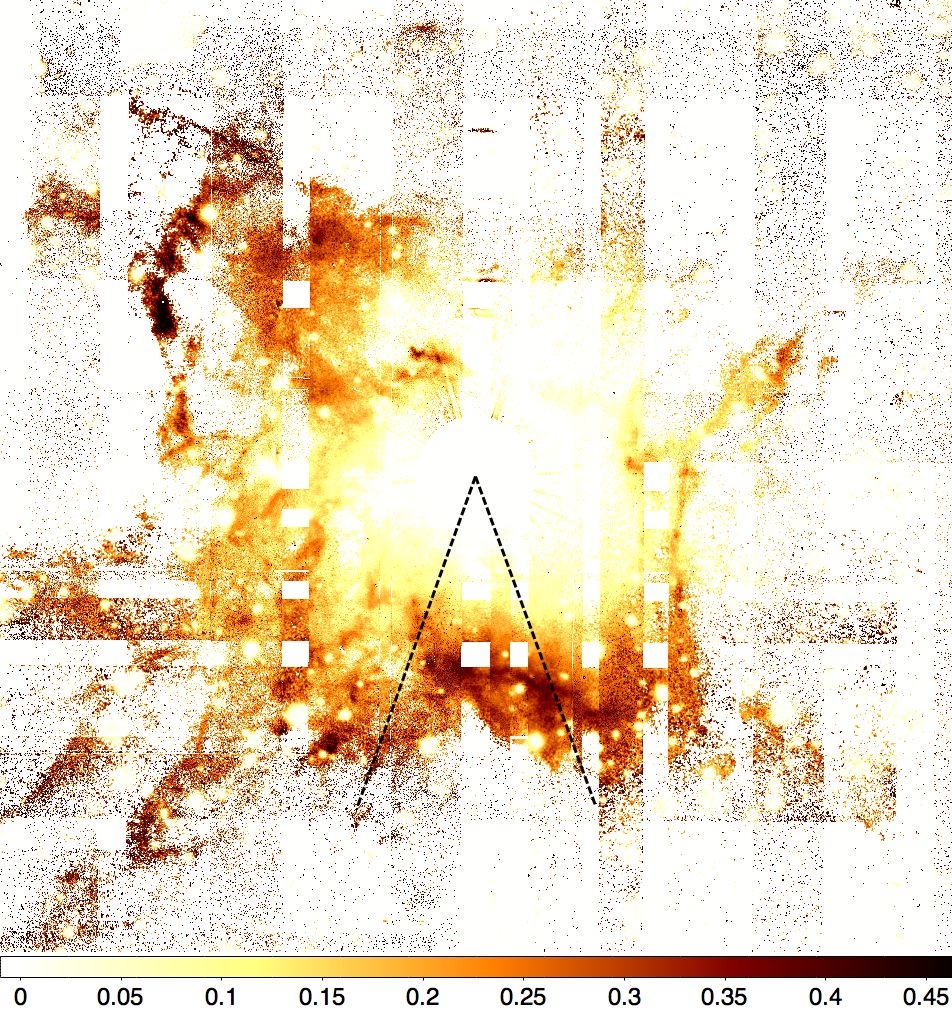}\hspace{3mm}
\includegraphics[width=9cm]{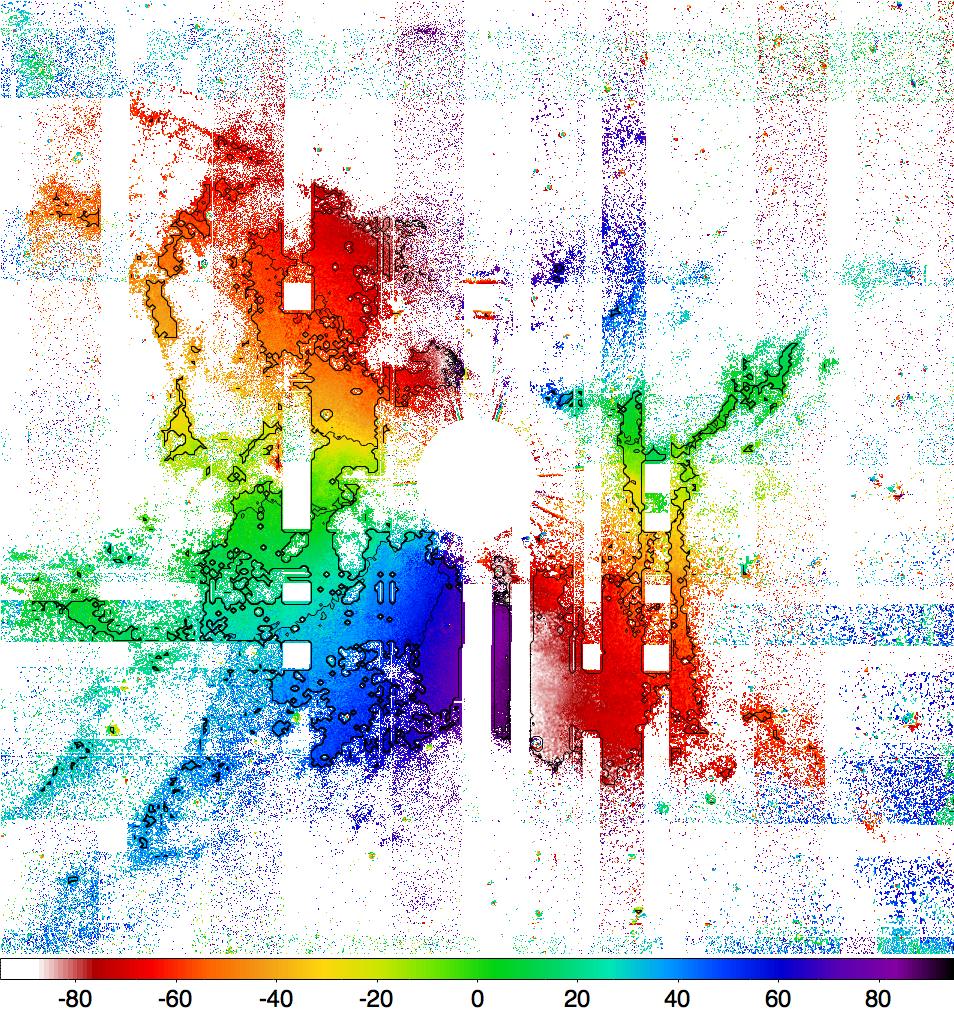}
\caption{{\it Left:} linear polarization degree $p_{\rm L}$ measured on the nebula of RS Pup. The dashed lines delimit the sector used to compute the profile presented in Fig.~\ref{radprof_pl}. {\it Right:} position angle $\theta$ of the polarization vectors {\bf (in degrees), with contour plots from -90 to +90$^\circ$ (steps of 30$^\circ$).} The $p_{\rm L}$ and $\theta$ maps result from the combination of FORS epoch 2 (reference) and epoch 1 (to fill the missing vertical bands). The field of view is $4\arcmin \times 4\arcmin$, with North up and East to the left in both cases.\label{PL_combined}}
\end{figure*}

\begin{figure}[ht]
\centering
\includegraphics[width=8cm]{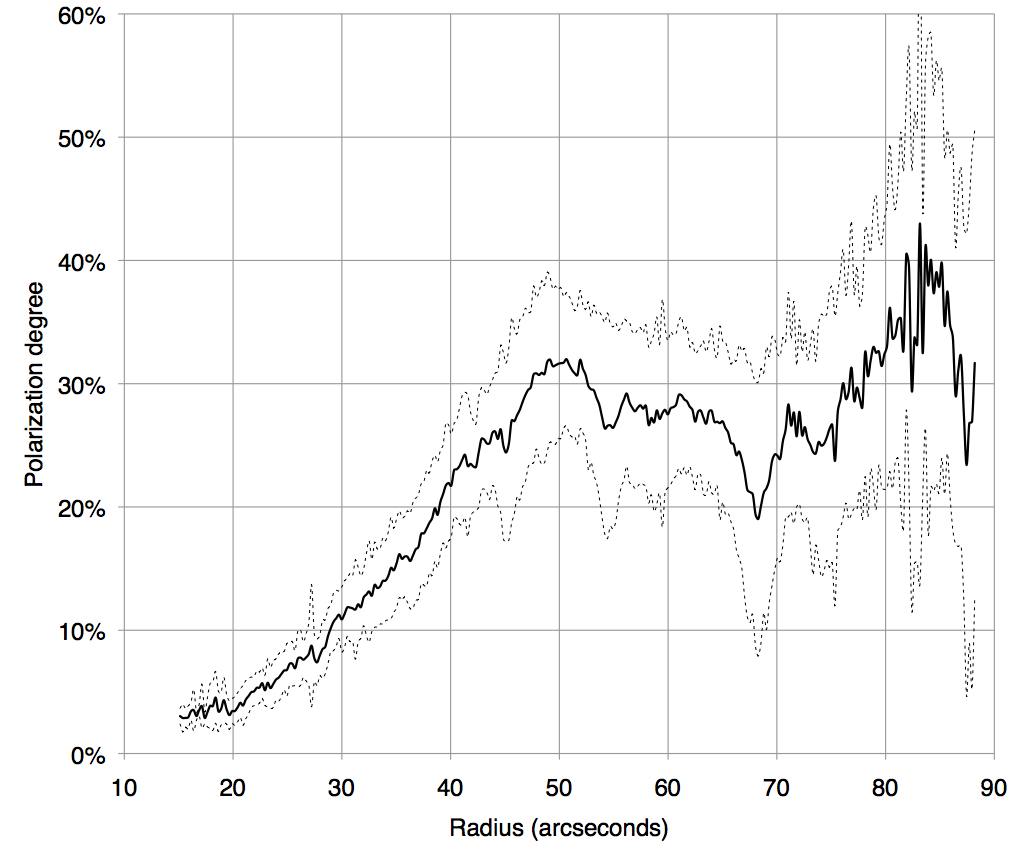}
\caption{Radial profile of the polarization degree $p_{\rm L}$ towards the south of RS\,Pup, as a function of the angular radius from the star. \label{radprof_pl}}
\end{figure}

\begin{figure}[ht]
\centering
% 952 pixel images = 4.0' field
\includegraphics[width=4.4cm]{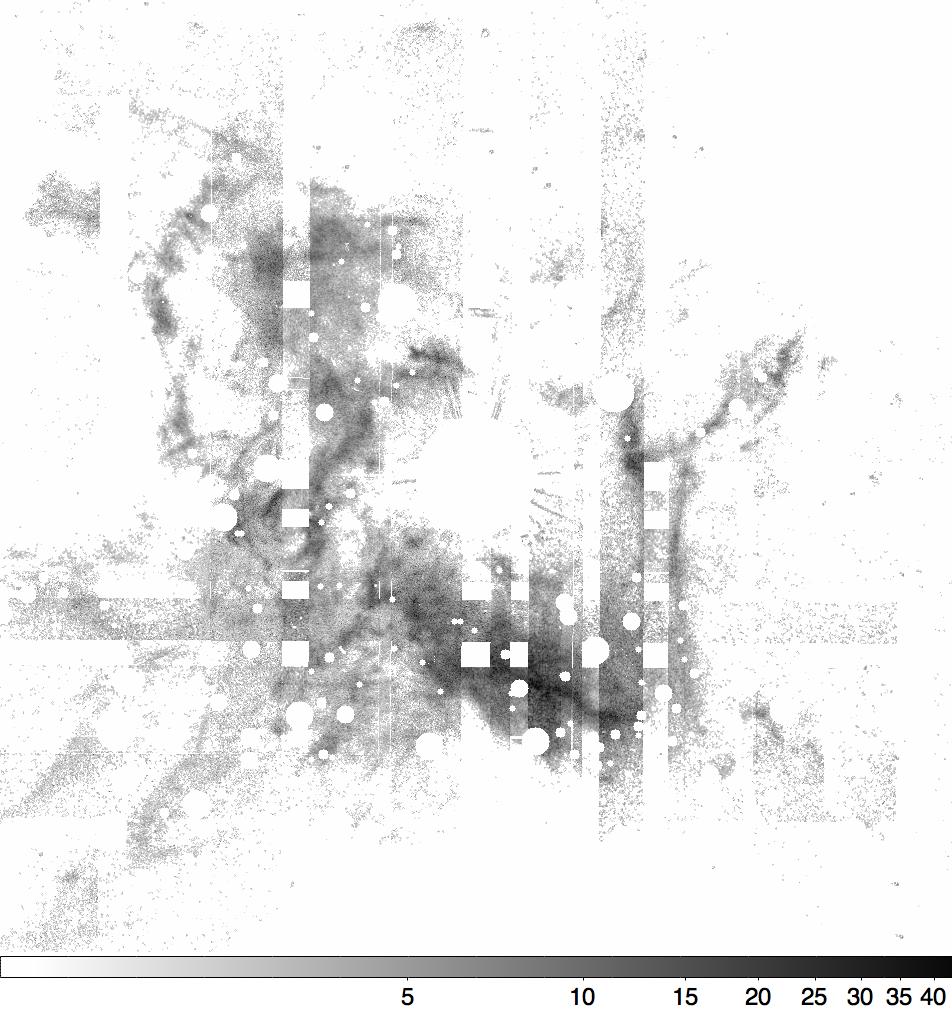}
\includegraphics[width=4.4cm]{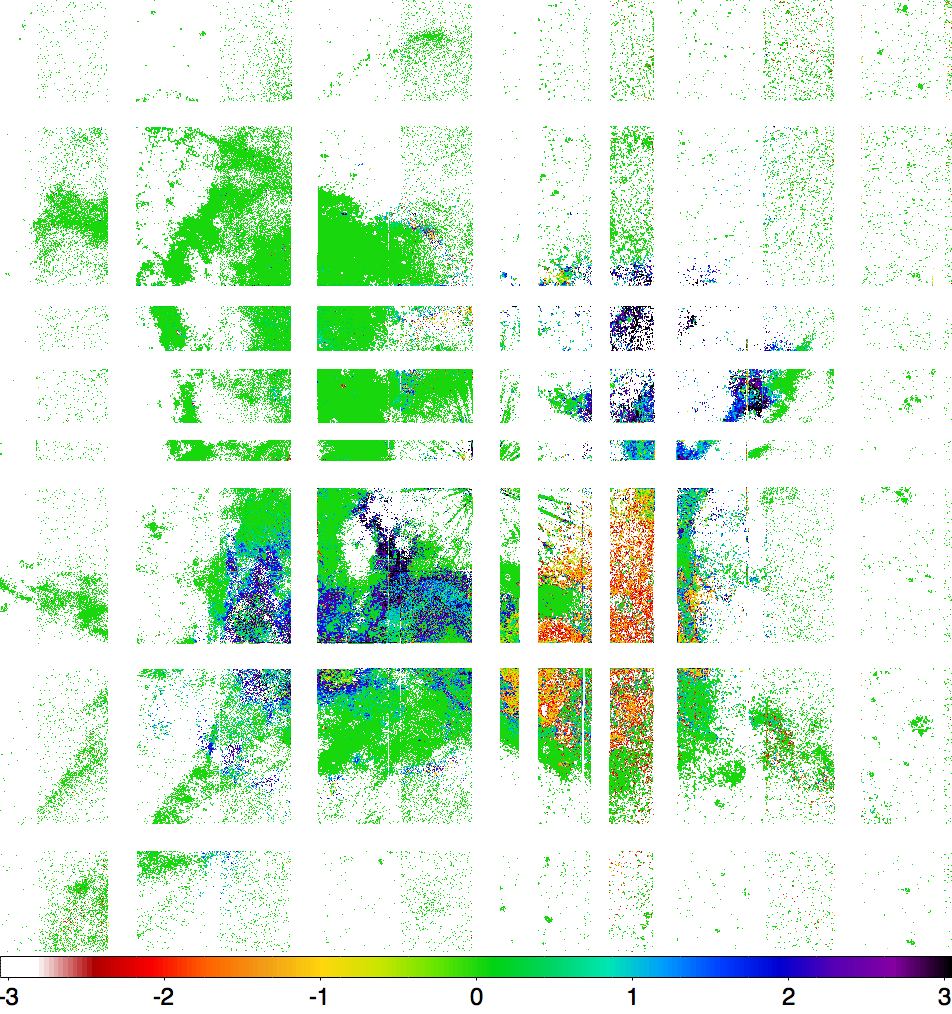}
\caption{{\it Left:} Map of the Signal to Noise Ratio (SNR) of the linear polarization degree $p_{\rm L}$. The intensity scale is logarithmic.
{\it Right:} Difference in measured polarization degree $p_{\rm L} ({\rm epoch 2}) - p_{\rm L} ({\rm epoch 1})$, normalized to the combined local standard deviation of the two images. The field of view is $4\arcmin \times 4\arcmin$, with North up and East to the left.
\label{PL_SNR_diff}}
\end{figure}

We precisely co-aligned our two observation epochs using the background stars as astrometric references.
We considered epoch 2 as the reference map, and we used epoch 1 only to fill the missing parts of the $p_{\rm L}$ and $\theta$ maps.  A moving average scaling coefficient was applied to the $p_{\rm L}$, $\theta$ and intensity maps of epoch 1 so that they match the local average value of epoch 2. The moving average normalization was computed over a square area of 60\,pixels ($15\arcsec$), and applied multiplicatively to each pixel of the epoch 1 image. This procedure ensures a smooth transition of the two images and reduces the artefacts. We therefore did not average the values obtained at the two epochs for each point on the nebula, even where they overlap. There are three reasons for this choice:
\begin{itemize}
\item The subtraction of the instrumental PSF wing halo is significantly more efficient for epoch 2 as for epoch 1, as the PSF calibrator image was obtained immediately after the epoch 2 on RS\,Pup, under identical seeing conditions, and with the instrument rotator positioned the same way. The subtraction of the PSF wings for epoch 1 (obtained 7 weeks earlier) was done using this same calibrator observation, therefore less efficiently.
\item Epoch 2 is of significantly better overall quality than epoch 1, due mainly to a lower sky background, and also to slightly better seeing conditions (see Table~\ref{fors_log}). The mismatch in photometric depth and in angular resolution between the two epochs would result in artefacts on the combined $p_{\rm L}$ and $\theta$ map.
\item The averaging of $p_{\rm L}$ and $\theta$ values is complicated by their peculiar statistical properties, and the potential resulting biases would be difficult to assess.
\end{itemize}
The resulting $p_{\rm L}$ and $\theta$ maps are presented in Fig.~\ref{PL_combined}, and the map of the Signal to Noise Ratio (SNR) $p_{\rm L}/\sigma(p_{\rm L})$ is shown in the left panel of Fig.~\ref{PL_SNR_diff}. The typical SNR on the intermediate surface brightness parts of the nebula is between 5 and 10 per pixel, and reaches more than 50 per pixel on the ridge-like structure located south of RS\,Pup (see also Sect.~\ref{scatter}). A representative profile of $p_{\rm L}$ is presented in Fig.~\ref{radprof_pl}. This radial curve was computed by averaging the measured $p_{\rm L}$ values between position angles 160$^\circ$ and 200$^\circ$ (measured from North). The standard deviation of the $p_{\rm L}$ values for each radius over this sector is shown with dashed lines.

%______________ Figure

It is interesting to remark that the position angle of the polarization vector is always perpendicular to the dust-star direction over the whole surface of the nebula. Such a behavior is caused by the simple geometry of the system, where the central Cepheid illuminates very small dust particles from a single point. With no particular microscopic orientation of the particles (that could be caused by magnetic fields for example), this creates purely Rayleigh scattering in central symmetry. As the position angle of the polarization vector does not constrain the nebula geometry, we will not use it for the rest of our analysis.

%______________ Figure

The difference between the $p_{\rm L}$ maps at the two epochs (epoch 2 $-$ epoch 1) is presented in Fig.~\ref{PL_SNR_diff} (right panel).
The linear intensity scale of this image gives the difference between the two epochs for each point in the nebula, expressed in number of times the combined standard deviation. The average difference between the two images is +0.4$\sigma$. The overestimation of the polarization degree at epoch~1 in the southwest quadrant of the nebula (red region in the right panel of Fig.~\ref{PL_SNR_diff}) and its underestimation along a southeast-northwest axis (blue regions) is likely due to the less efficient asymmetric PSF halo subtraction for this epoch (exposures \#1-8). The slightly higher polarization degree found at epoch~2 over the rest of the nebula is most probably the result of the better seeing conditions and the significantly lower sky background.

%______________ Figure
\begin{figure}[t]
\centering
% 952 pixel images = 4.0' field
\includegraphics[width=\hsize]{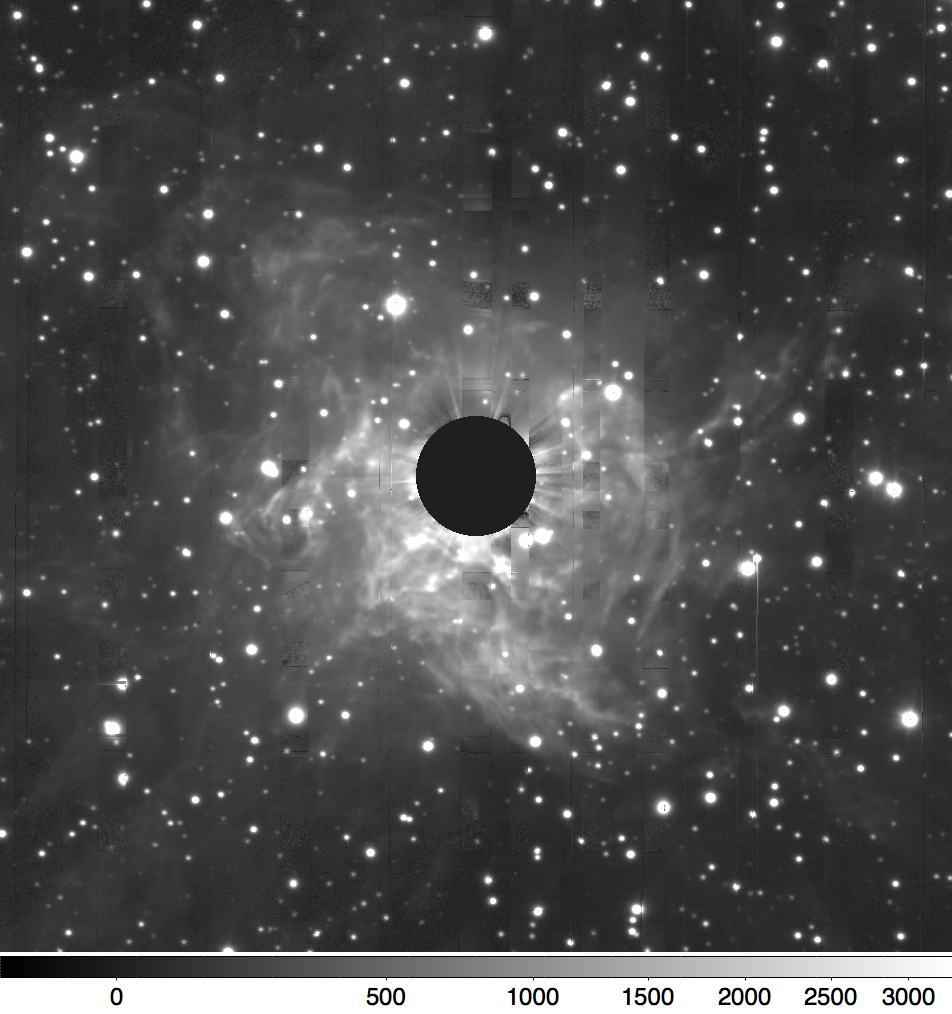}
\caption{Total scattered light intensity (in FORS ADUs) from the combination of FORS epochs 1 and 2, as well as an EMMI $V$ band image to cover the remaining gaps (see text for details). The field of view is $4.3 \arcmin \times 4.3\arcmin$.\label{Intensity_combined}}
\end{figure}

The combined intensity image from epochs 1 and 2 still contains small square unobserved patches, at the locations where the FORS stripe masks of the two epochs cross each other. To retrieve the intensity of the nebula on these patches, we used an EMMI observation obtained on 27 January 2007 in the $V$ band (Fig.~\ref{EMMI8770}, see also Paper~I for details).
The phase of this image is not identical to the phases of the FORS epochs 1 and 2, but was obtained at phase 0.877, i.e. approximately in the middle of the two FORS phases. As the epochs do not all correspond to the same phase of the Cepheid, the intensity distribution of the light echoes on the nebula is not strictly identical. This results in a slight mismatch of some sections of the nebula in intensity, and to discontinuities in the combined intensity image. However, these discontinuities remain at a reasonably low level and do not affect significantly the aspect of the nebula. The combined intensity image is presented in Fig.~\ref{Intensity_combined}.

\subsection{Interpolation of the $p_{\rm L}$ map}

\begin{figure}[t]
\centering
% 952 pixel images = 4.0' field
\includegraphics[width=\hsize]{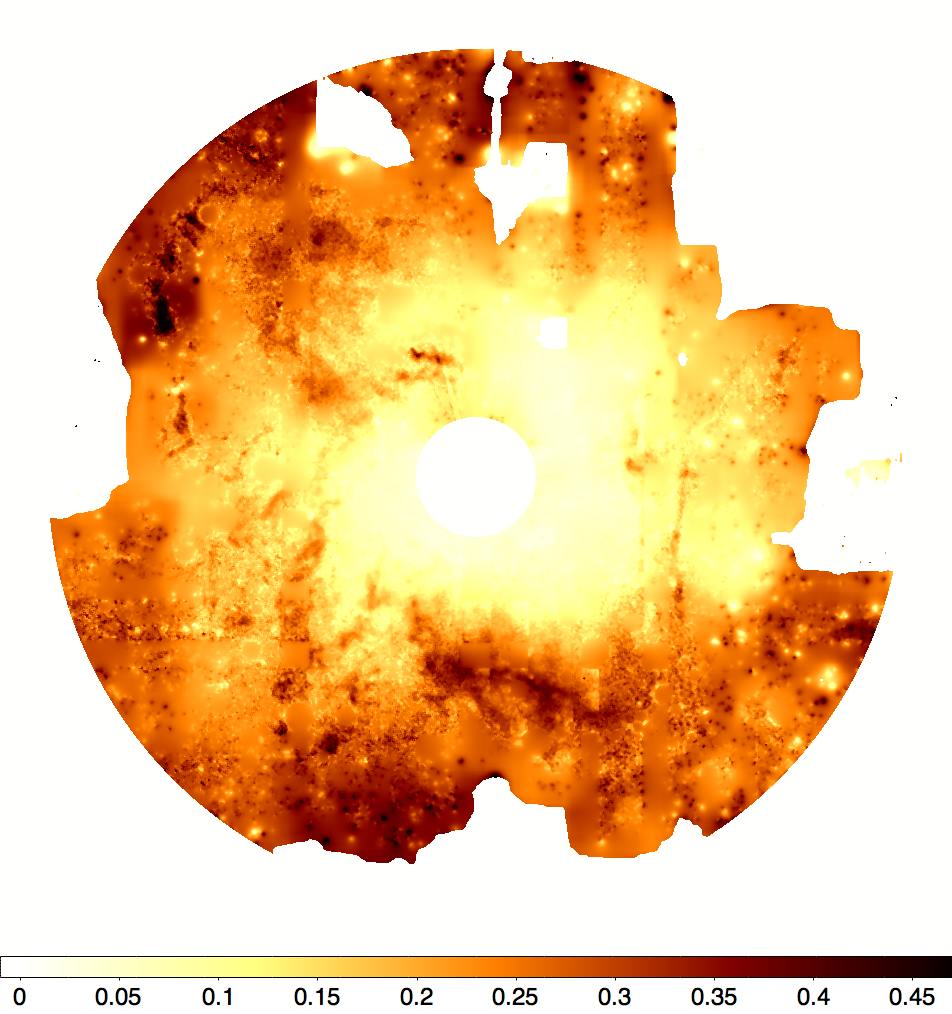}
\caption{Interpolated map of the inear polarization degree $p_{\rm L}$ over the RS\,Pup nebula (from Fig.~\ref{PL_combined}, left panel). The field of view is $4.3 \arcmin \times 4.3\arcmin$.\label{PL_combined_interp}}
\end{figure}

The accurate retrieval of the 3D structure and mass of the nebula requires that the $p_{\rm L}$ map has a coverage as continuous as possible of the nebula. Due to the low surface brightness of the outer extensions of the nebula, as well as the presence of gaps due to our observing strategy, we therefore need to interpolate the $p_{\rm L}$ values over the nebula. The missing values were computed using a moving average over a 60\,pixels square box. The pixels where $p_{\rm L}$ is available within this box were weighted by the square of their distances to the interpolated pixel. The field stars visible in the image were masked before this operation to avoid biases on the degree of linear polarization due to the star haloes. The combined $p_{\rm L}$ image was then median filtered over a $3\times3$ box to reduce the apparent pixel-to-pixel noise, and the resulting map is presented in Fig.~\ref{PL_combined_interp}.

%______________ Figure

%__________________________________EMMI Observations
\section{Light echo contrast from NTT/EMMI observations\label{emmiobs}}

The observations discussed in this Section were already presented in details in Paper~I. The processed image data set consists of a series of six $B$ band images of the nebula from which the instrumental/atmospheric scattered light halo of the central star was subtracted carefully using a PSF reference. These images (Fig.~6 in Paper~I) show clearly the propagation of the light echoes over the apparent surface of the nebula.

The fit of the photometric variation curve of RS\,Pup was obtained by adjusting the phase $\phi(\alpha,\delta)$, mean value $\overline{I}(\alpha,\delta)$ and amplitude $\Delta I(\alpha,\delta)$ of the Cepheid photometric variation curve to the flux measured at each position $(\alpha,\delta)$ in the image (where the SNR is sufficient). We used a classical $\chi^2$ minimization through a Levenberg-Marquardt algorithm. The fit was obtained separately on each pixel of $5\times5$~pixel boxes.
The standard deviation of the measurements over the $5\times5$~pixel box, added quadratically to the statistical fitting error bars, was taken as its associated uncertainty. While the estimation of each phase offset is more precise, this conservative approach is intended to account for possible photometric contaminations by other nearby features.

\begin{figure}[ht]
\centering
\includegraphics[width=\hsize]{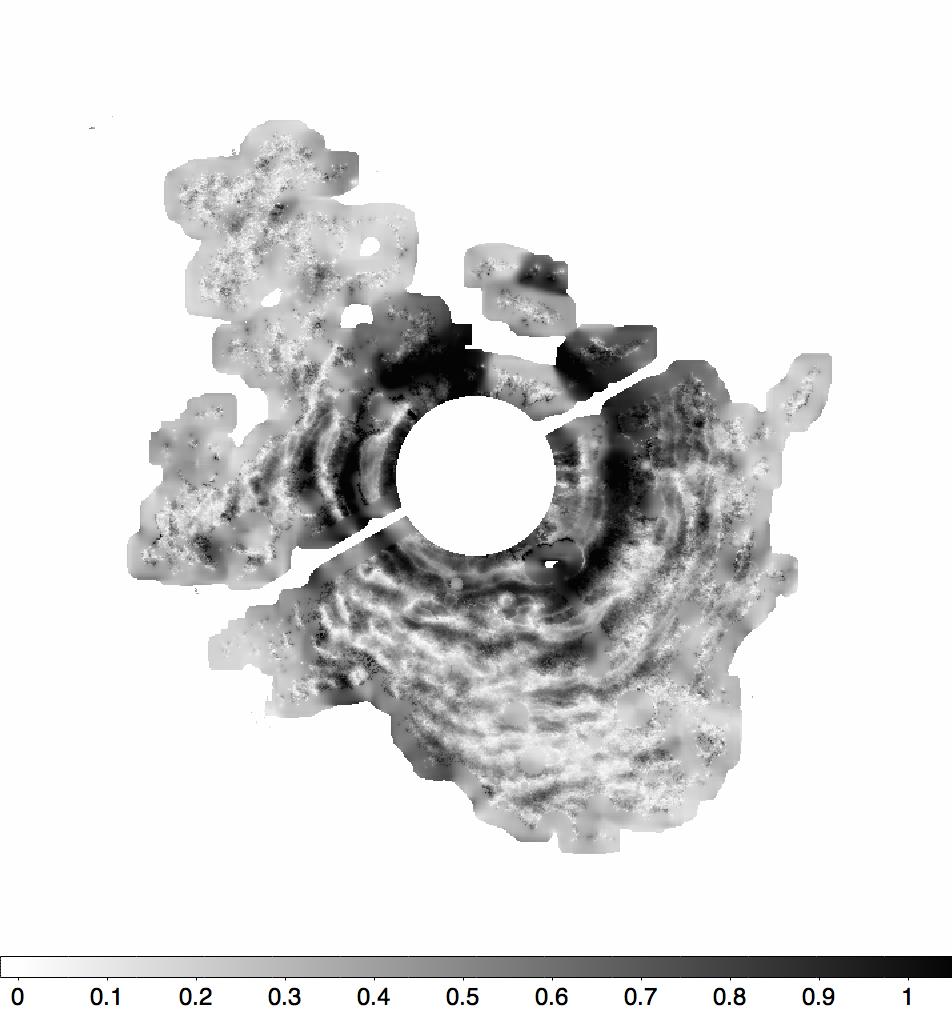}
\caption{Map of the light echoes normalized contrast $C(\alpha,\delta)$ in the nebula of RS\,Pup. The ring-like structure close to the star is probably an artefact caused by the limited phase coverage of our EMMI images (see text). The field of view is $3\arcmin \times 3\arcmin$, with North up and East to the left. \label{contrast_map}}
\end{figure}

\begin{figure}[ht]
\centering
\includegraphics[width=8cm]{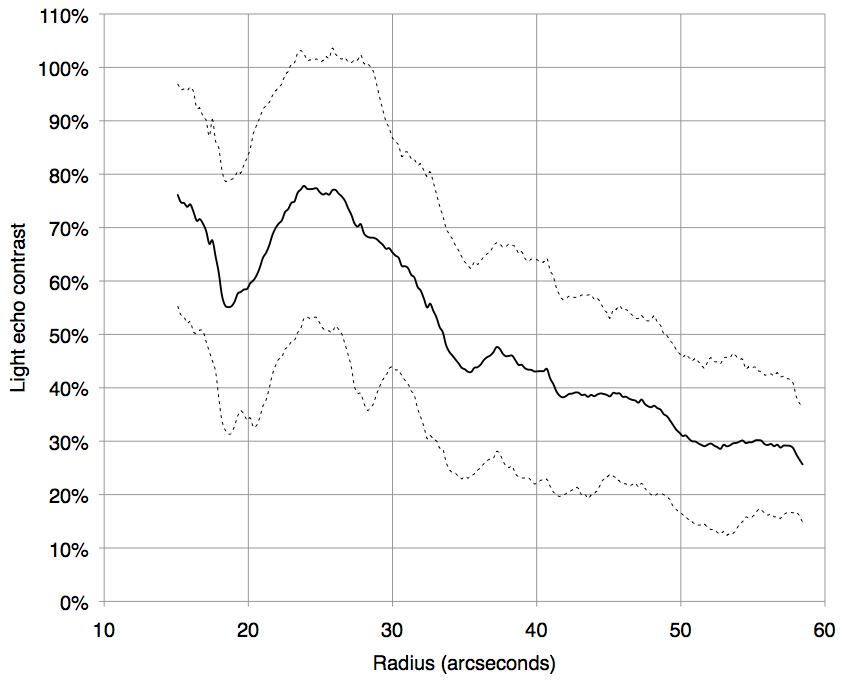}
\caption{Azimuthally averaged radial profile of the light echo contrast $C$ on the nebula of RS\,Pup, as a function of the angular radius from the star. The dashed curves show the boundaries of the $\pm\sigma$ domain. \label{radprof_contrast}}
\end{figure}

As a result, we obtain a map of the normalized contrast $C(\alpha,\delta)$ of the photometric variation over the nebula, relative to the contrast of the photometric light curve of RS\,Pup in the $B$ band $C_0$:
\begin{equation}
C(\alpha,\delta) = \frac{\Delta I(\alpha,\delta)}{\overline{I}(\alpha,\delta)} \, \frac{1}{C_0}
\end{equation}
with $C_0 = {\Delta I_\mathrm{RS\,Pup}} / {\overline{I_\mathrm{RS\,Pup}}} = 1.69$.
The resulting normalized contrast map is presented in Fig.~\ref{contrast_map}, and an azimuthally averaged radial profile is shown in Fig.~\ref{radprof_contrast}.
This map presents rings close to the star. The probable reason of their presence is that the phase coverage of our EMMI CCD images (Paper~I) is not sufficient to have a good 3-parameter fit (phase/amplitude/average value) at all locations over the nebula. This is visible in Fig. 8 of Paper~I, as most of the measurement points (6) were obtained between phases 0.5 and 1.0, and only one point is present between phases 0 and 0.5. At some locations of the nebula, we therefore do not sample properly the maximum and minimum flux phases, which makes the fit more difficult. In this case, the amplitude is likely underestimated, although the phase is probably much less affected. We use this map in Sect.~\ref{nebulathickness} to estimate the thickness of the light scattering layer. 
The phase map will be used in a forthcoming article to estimate the geometrical distance of RS\,Pup.

%______________ Figure

%______________ Figure

%__________________________________Model
\section{Properties of the light-scattering nebula\label{3Dshape}}

\subsection{Thickness of the scattering layer\label{nebulathickness}}

From the observed amplitude of the light echoes $C(\alpha,\delta)$ (Sect.~\ref{emmiobs}), it is possible to estimate the thickness of the scattering layer. A very thin veil of dust will result in a perfect contrast of the echoes compared to the star's light curve. A thick dust layer will naturally decrease the echoes contrast, as a continuum of light-scattering particles with different phases are then present on the line of sight, resulting in a smeared light curve (Sugerman~\cite{sugerman03}, Bond \& Sparks~\cite{bond09}). Qualitatively, the fact that we observe contrasted light echoes indicates that the dust is spread over a thin surface, compared to the spatial extent of a propagating echo $c \times P$ (with $c$ the speed of light and $P$ the period of the Cepheid), rather than within a thick layer.

\begin{figure}[ht]
\centering
\includegraphics[width=8cm]{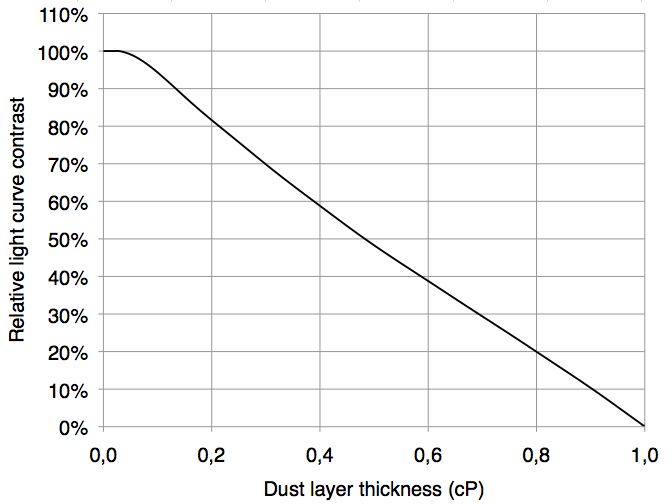}
\caption{Model light echo normalized contrast $C(e)$ (see Eq.~\ref{eq:contrast}) as a function of the equivalent thickness of the scattering layer, expressed in units of $c \times P = 7180$\,AU.\label{softening_curve}}
\end{figure}

Following a similar method as Havlen~(\cite{havlen72}), we define a range of softened light curves $I_s(\phi, e)$ for different dust layer thicknesses $e$ by convolving the Cepheid light curve $I(\phi)$ (with $\phi$ the phase) in the $B$ band with a square function of variable width $e$. The thickness $e$ of the considered homogeneous dust layer model is expressed in units of $c \times P = 7180$\,AU. We define the normalized contrast function $C(e)$ as:
\begin{equation}
%C(e) = \frac{\max \left[ I_s(\phi,e)\right] - \min \left[ I_s(\phi,e)\right] }{\max \left[ I_s(\phi,e)\right]  + \min \left[ I_s(\phi,e)\right] }
C(e) = \frac{\Delta I_s(e) }{ \overline{I_s}(e) } \, \frac{1}{C_0}
\label{eq:contrast}
\end{equation}
where $\Delta I_s$ is the amplitude of the softened photometric light curve, $\overline{I_s}(e)$ its average value, and $C_0=1.69$ the contrast of the light curve of RS\,Pup in the $B$ band.
The shape of the $C(e)$ function is shown in Fig.~\ref{softening_curve}. Due to the sharp maximum light peak in RS\,Pup's light curve, this function is close to linear. Inverting numerically $C(e)$ gives the relation $e(C)$ between the thickness of the scattering layer and the observed contrast.

%______________ Figure

We computed the light echo contrast $C(\alpha, \delta)$ from the 3-parameter light curve fit on the EMMI image cube (Sect.~\ref{emmiobs}). We consider the parts of the nebula that are relatively close to the star, as we aim at small scattering angles. Large scattering angles, that would occur e.g. close to the edge of a spherical nebula, would result in an apparently higher thickness, due to projection effects. In order to derive the true thickness of the dust layer independently of projection effects, we thus focus our evaluation on the nebula within a radius of $\approx 30\arcsec$ from the star. 

%______________ Figure
\begin{figure}[ht]
\centering
\includegraphics[width=8cm]{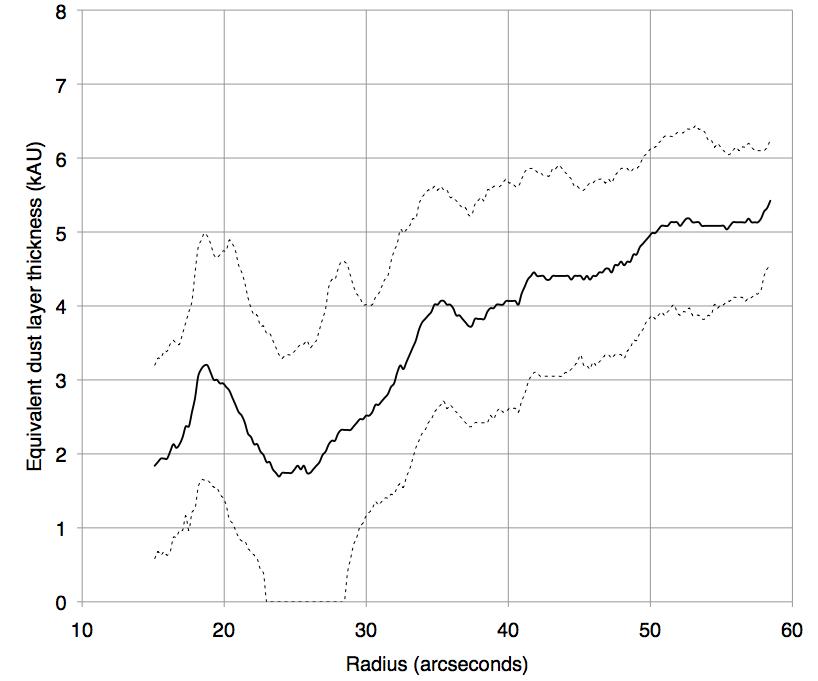}
\caption{Dependence of the equivalent geometrical thickness $e$ of the light scattering dust layer as a function of the angular radius from RS\,Pup (expressed in thousands of AU). The dashed curves indicate the boundaries of the $\pm \sigma$ domain.\label{thickness_curve}}
\end{figure}

From the observed contrast $C(\alpha, \delta)$ and model thickness $e(C)$, we derived the equivalent geometrical thickness $e$ of the scattering layer. Fig.~\ref{thickness_curve} shows the resulting thickness as a function of the angular radius from the star. We obtain a typical thickness of the dust layer of $e \approx 2 \pm 2$\,kAU over the considered area (within $30\arcsec$ from the star). At the expected distance of RS\,Pup ($1.8 \pm 0.1$\,kpc, from Fouqu\'e et al.~\cite{fouque07}, see also references therein), this corresponds to a relative thickness of only $\approx 1\%$ compared to the typical radius of the visible nebula (that extends on $\gtrsim 2\arcmin$, Sect.~\ref{nebshape}). The weakly significant increase of the derived thickness with the angular radius from RS\,Pup could be either due to a geometrical projection effect, or to a physical property of the nebula. It is however difficult to conclude firmly on this question, due to the limited accuracy of the measurement and the fact that this curve was computed through an azimuthal integration over the irregular nebula surface.

We conclude that the geometry of the nebula is essentially a single, geometrically and optically thin, light-scattering dust layer. Moreover, as discussed in Sect.~\ref{scatter}, forward scattering is far more efficient than backwards scattering. This means that the dust layer we observe in our EMMI images is essentially located between us and RS\,Pup, rather than behind the star. In the following, we thus adopt the model of a thin veil of dust located between us and the Cepheid to reproduce our observations.
Thanks to the high light echo contrast $C(\alpha,\delta)$ observed over the nebula, we neglect the ring artefacts mentioned in Sect.~\ref{emmiobs} (that are also visible in the azimuth averaged curve in Fig.~\ref{thickness_curve}) in this analysis, as the corresponding underestimation of the contrast does not affect our conclusion that the scattering dust layer is geometrically very thin.

\subsection{Polarization model\label{polar}}

The dependence of the linear polarization degree $p_{\rm L}$ on the scattering angle $\theta$ may be assumed to be given by the classical polarization phase function for Rayleigh scattering (White~\cite{white79}):
\begin{equation}\label{pl_eq}
p_{\rm L} = p_{\rm max}\,\frac{1-\cos^2 \theta}{1+\cos^2 \theta}
\end{equation}
where $p_{\rm max}$ is equal to 1 for theoretical Rayleigh scattering, but is lower in real astrophysical situations. The $p_{\rm L}(\theta)$ function reaches its maximum for $\theta = 90^\circ$, i.e. dust located in the plane of the sky.

\begin{figure}[ht]
\centering
\includegraphics[width=8cm]{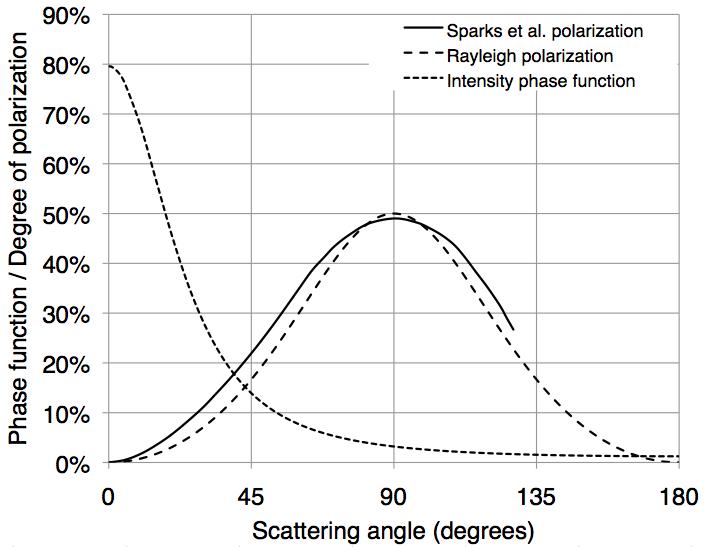}
\caption{Degree of linear polarization $p_{\rm L}(\theta)$ as a function of the scattering angle $\theta$ (in degrees). The empirical $p_{\rm L}(\theta)$ function from Sparks et al.~(\cite{sparks08}) is shown as a solid curve (extrapolated to small scattering angles), together with a classical Rayleigh scattering function with $p_{\rm max} = 0.50$ (dashed curve). The photometric Henyey-Greenstein phase function $\Phi(\theta)$ is represented for $g=0.538$ using a dotted curve.
\label{scatter_model}}
\end{figure}

Considering a maximum polarization value of $p_{\rm max} = 0.50 \pm 0.05$, Sparks et al.~(\cite{sparks08}) showed that the Rayleigh scattering model is a relatively good match to the observed polarization profile of the light echo of V838 Mon. But these authors also deduced from their observations an empirical polarization phase function (their Fig.~15) that gives a better match to their observations and slightly departs from the Rayleigh function (particularly for intermediate polarization degrees). The good reproducibility of this phase function for the different observation epochs of V838 Mon (i.e. different echo distances from the star) leads us to retain this empirical model for our analysis of RS\,Pup's polarization profile.
The dependence of the degree of linear polarization as a function of the scattering angle is presented in Fig.~\ref{scatter_model}.
Inverting numerically this function gives us access to the scattering angle $\theta$, under the assumption that it is lower than $90^\circ$. This hypothesis is justified by the fact that backward scattering is much less efficient than forward scattering, as shown by the shape of the photometric Henyey-Greenstein phase function $\Phi(\theta)$ represented in Fig.~\ref{scatter_model} (see also Sect.~\ref{scatter}). As a consequence, we observe essentially the light scattered by the dust located between us and RS\,Pup (as opposed to the dust located behind the star). We may observe particularly dense dust features located slightly beyond the plane of the sky (i.e. with scattering angles $\theta > 90^\circ$), but the scattered light flux will be low compared to forward scattering features.

%______________ Figure

\subsection{Three-dimensional dust distribution\label{3Ddust}}

%______________ Figure
\begin{figure}[ht]
\centering
\includegraphics[width=7cm]{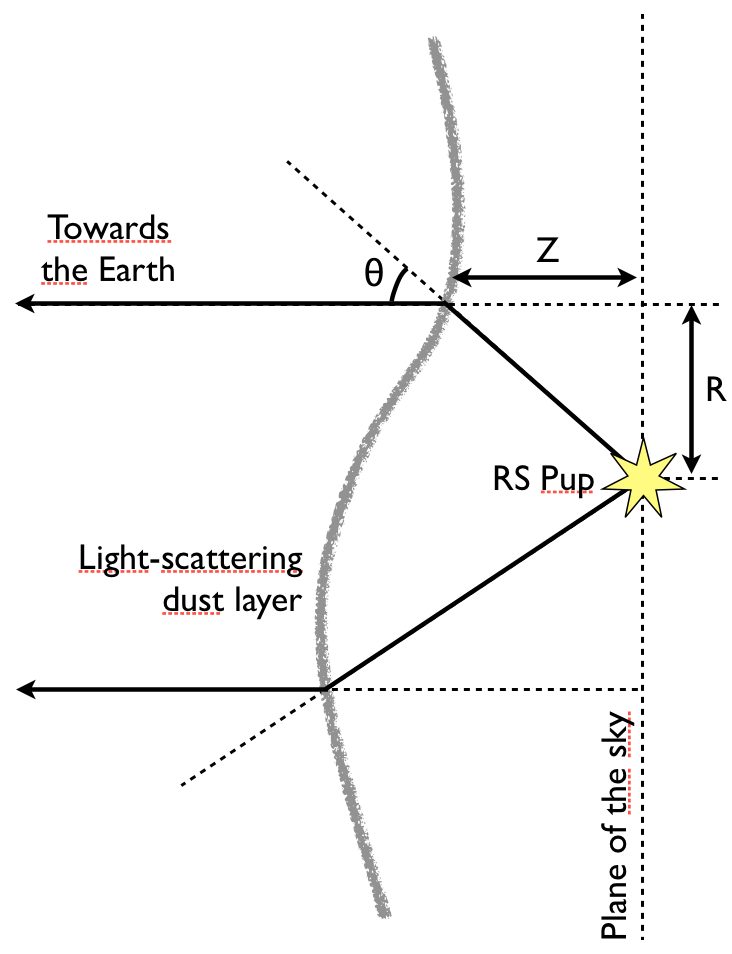}
\caption{Geometry of the adopted scattering model.\label{geometry}}
\end{figure}

\begin{figure*}[ht]
\sidecaption
% 952 pixel image, field of 4', horizontal color scale, number font size = 24
\includegraphics[width=12cm]{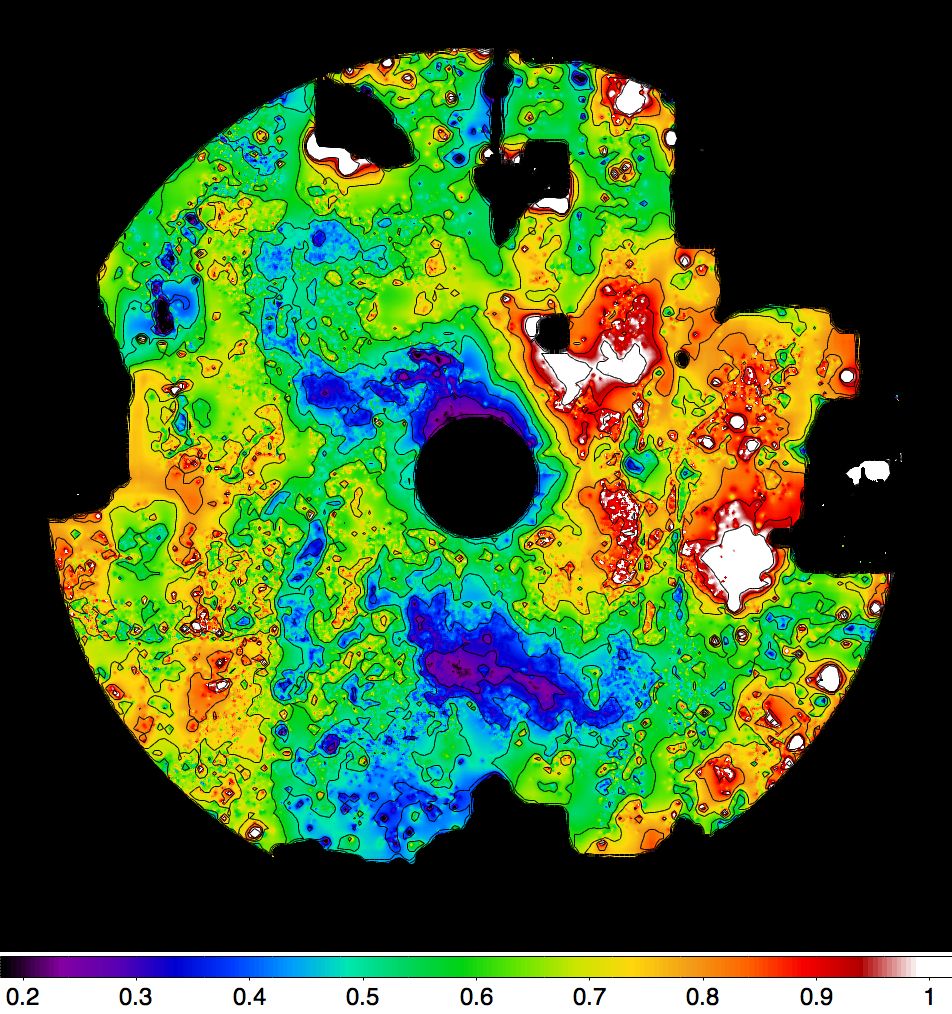}
\caption{Altitude of the light scattering dust layer relative to the plane of the sky at RS\,Pup's distance (assumed to be $1.8 \pm 0.1$\,kpc), expressed in parsecs, with contour curves spaced by 0.1\,pc. The field of view is $4\arcmin \times 4\arcmin$, equivalent to $2.1 \times 2.1$\,pc at the distance of RS\,Pup, with North up and East to the left. The nebular features located further than $1.8\arcmin$ from RS\,Pup are masked (too low SNR).  \label{Zimage_pc}}
\end{figure*}

Under the assumption that we see only forward scattering dust ($\theta < 90^\circ$) it is possible to invert numerically the polarization phase function from Sparks et al.~(\cite{sparks08}) shown in Fig.~\ref{scatter_model} to retrieve the scattering angle $\theta$. The overall geometry of the adopted single layer scattering model is presented in Fig.~\ref{geometry}.
The projected distance $R$ of each point of the nebula from RS\,Pup is directly measurable on the image, and we can therefore deduce $Z$ immediately, knowing $\theta$, from:
\begin{equation}
Z (\alpha,\delta) = \frac{R (\alpha,\delta)}{\tan \left[ \theta(\alpha,\delta) \right] }
\end{equation}
In this expression, $Z$ has the same physical unit as $R$, and it can be expressed either as an angle (expressed, e.g. in arcseconds), or as a linear distance if we consider the distance to RS\,Pup known {\it a priori}. We chose this second approach, assuming a distance of $1.8 \pm 0.1$\,kpc for RS\,Pup ($\pi = 0.55 \pm 0.03$\,mas, from Fouqu\'e et al.~\cite{fouque07}). The pixel scale of our FORS polarimetric data ($0.252\arcsec$/pix) thus translates into a linear scale of $2.22 \times 10^{-3}\,\mathrm{pc/pix}=458\,\mathrm{AU/pix}$ at the distance of RS\,Pup. The resulting ``altitude" of the dust layer above the plane of the sky at RS\,Pup's distance is shown in Fig.~\ref{Zimage_pc}. The relative uncertainty on this altitude is estimated to $\approx 5$ to 15\% within $1\arcmin$ of the central star.

As a remark, the true distribution of the dust likely departs from our simple, single-layer model at some locations over the nebula. This will be the case for instance if there are several distinct layers of dust located on the line of sight. In this case, our altitude estimate corresponds to the mean of the altitudes of the different dust layers, weighted by the product of the dust density by the phase function $\Phi(\theta)$ of each contributing layer.

%______________ Figure

\subsection{Dust column density\label{scatter}}

To derive the scattered light intensity, we adopt the Henyey-Greenstein~(\cite{henyey41}) phase function (Fig.~\ref{scatter_model}):
\begin{equation}\label{phi_eq}
\Phi(\theta) = \frac{1-g^2}{4\pi\,(1+g^2-2g\, \cos \theta)^{3/2}}
\end{equation}
where $\theta$ is the scattering angle ($\theta = 0^\circ$ corresponds to forward scattering), and $g$ the scattering asymmetry factor for which we adopt a value of $g=0.538$ in the $V$ band (Draine~\cite{draine03a}, see also Draine~\cite{draine03b} and White~\cite{white79}).
Eq.~\ref{phi_eq} provides us with a way to measure the column density of hydrogen atoms $n_\mathrm{H}\,(\alpha,\delta)$ of the scattering material (in H\,arcsec$^{-2}$):
\begin{equation}\label{density_eq}
n_\mathrm{H}\,(\alpha,\delta) =\frac{\overline{I}(\alpha,\delta)}{\overline{F}}\,\frac{4\,\pi\,d^2}{\omega\,\Phi(\theta)\,\sigma}
\end{equation}
where $\overline{I}(\alpha,\delta)$ is the average surface brightness of the nebula as measured at the Earth at celestial coordinates $(\alpha,\delta)$ (in W\,m$^{-2}$\,$\mu$m\,arcsec$^2$), $d$ is the linear distance from RS\,Pup to the scattering material (in meters), $\overline{F}$ the apparent average flux of the star at the Earth (in W\,m$^{-2}$\,$\mu$m), $\omega$ the dust albedo, and $\sigma$ is the scattering cross section per H nucleon.
We adopt the case A PAH/graphite/silicate grain Milky Way model ($R_V=A_V/E(B-V)=3.1$) developed by Weingartner \& Draine~(\cite{weingartner01}), and described by Draine~(\cite{draine03a}): $\omega = 0.674$ and $\sigma = 3.275 \times 10^{-26}\,$m$^{2}/\mathrm{H}$ in the $V$ band\footnote{http://www.astro.princeton.edu/$\sim$draine/dust/scat.html}. Considering the dispersion of the different scattering models, we estimate that a reasonable range of values for the product $\omega \times \sigma$ is $\pm 15$\% (see e.g. Kr\"ugel~\cite{kruegel09}). This factor impacts linearly the determination of the density of the scattering nebula, and we adopt this range as our systematic uncertainty due to the chosen dust model.
%
%______________ Figure
\begin{figure}[]
\centering
\includegraphics[width=\hsize]{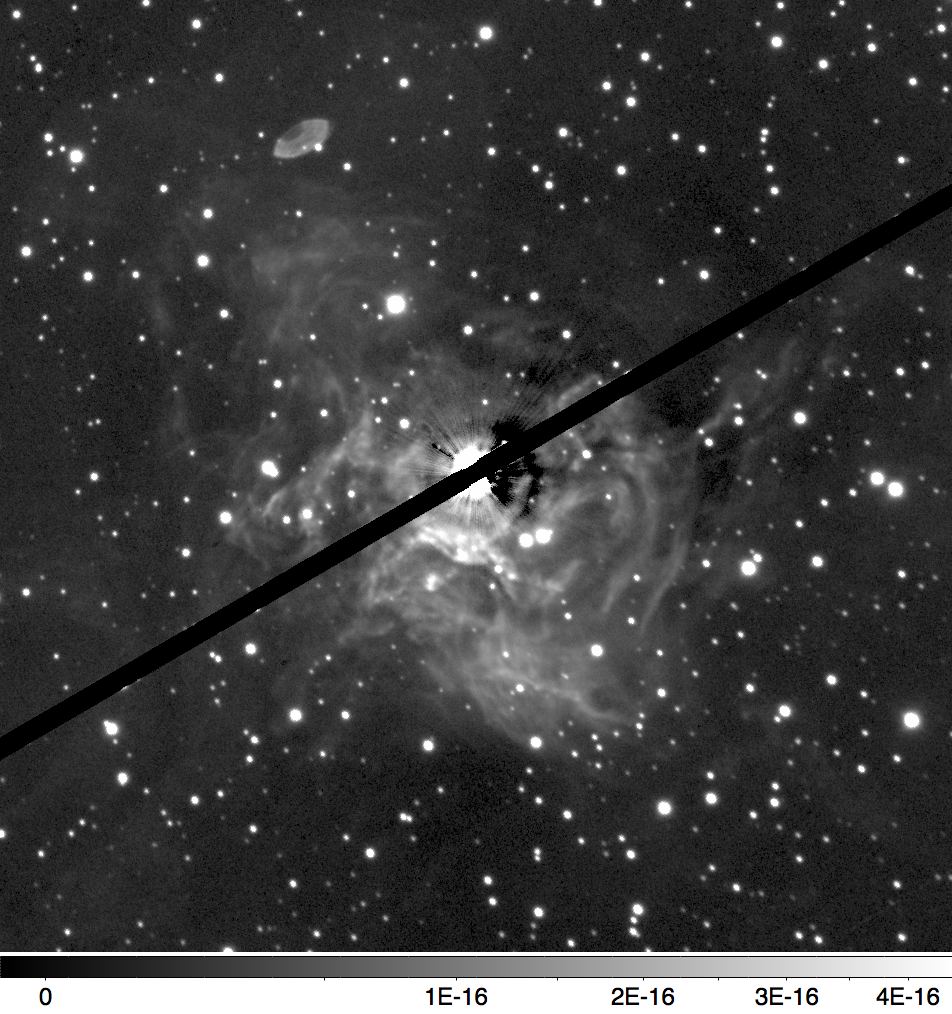}
\caption{EMMI image of the nebula of RS\,Pup in the $V$ band, obtained at pulsation phase $\phi=0.877$. The intensity scale follows a square root law, and is expressed in W\,m$^{-2}$\,$\mu$m$^{-1}$\,sr$^{-1}$. The field of view is $4\arcmin \times 4\arcmin$. \label{EMMI8770}}
\end{figure}
The only remaining variable to measure in Eq.~\ref{density_eq} to obtain $n_\mathrm{H}$ is the distance $d$ between the central star and the scattering material. We deduce this distance from the geometrical model of the nebula presented in Fig.~\ref{Zimage_pc}. For this work, we used the EMMI $V$ band image presented in Fig.~\ref{EMMI8770} (see also Paper~ÊI). We preferred this approach to using the FORS image presented in Fig.~\ref{PL_combined}, as the photometric calibration of that direct CCD image is more reliable than the polarimetric observations obtained with FORS. The resulting map of the column density $n_\mathrm{H}$ around RS\,Pup is presented in Fig.~\ref{nH_density}.

%______________ Figure
\begin{figure*}[ht]
\sidecaption
% 952 pixel image, field of 4', horizontal color scale, number font size = 24
\includegraphics[width=12cm]{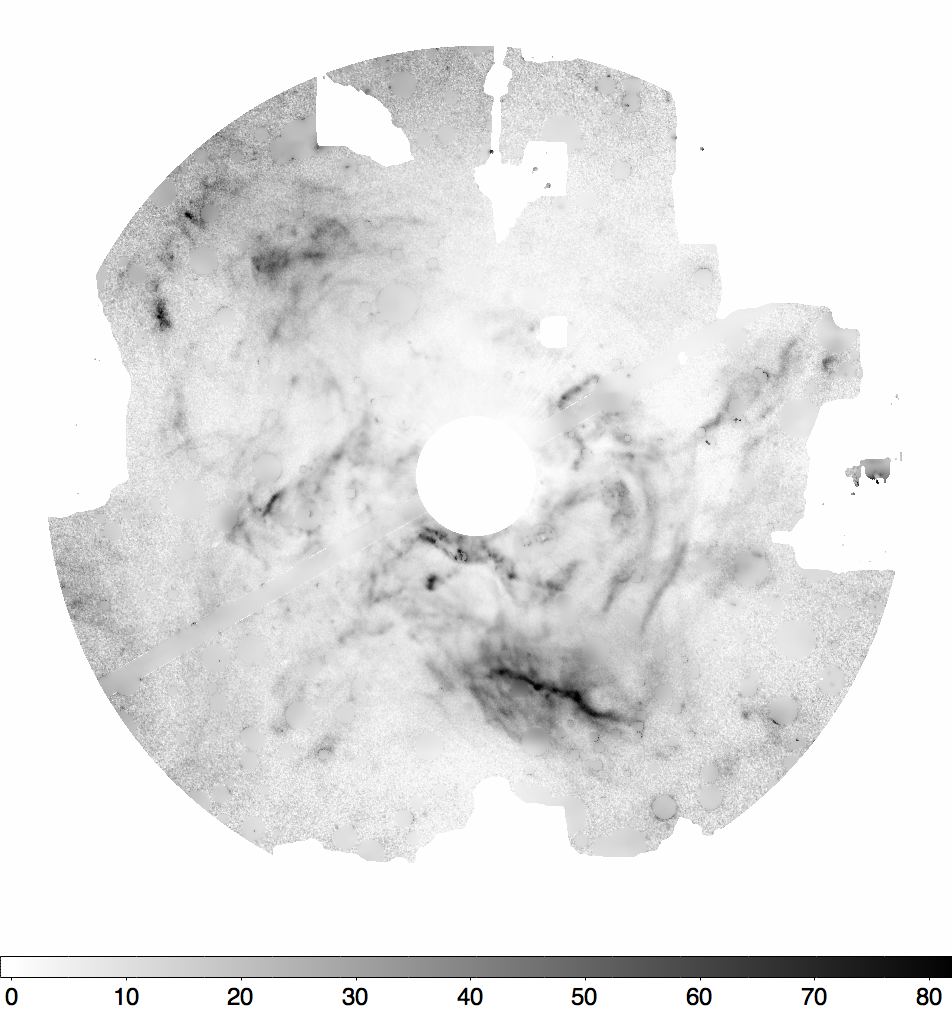}
\caption{Column density of hydrogen atoms $n_\mathrm{H}$ of the scattering material around RS\,Pup, in unit of $10^{54}\,\mathrm{H}\,\mathrm{arcsec}^{-2}$, limited to an angular radius of $1.8\arcmin$. The field stars have been removed from the image for clarity. The field of view is $4\arcmin \times 4\arcmin$, with North up and East to the left.\label{nH_density}}
\end{figure*}

\subsection{Dust geometry model validation}

\begin{figure}[]
\centering
\includegraphics[width=7cm]{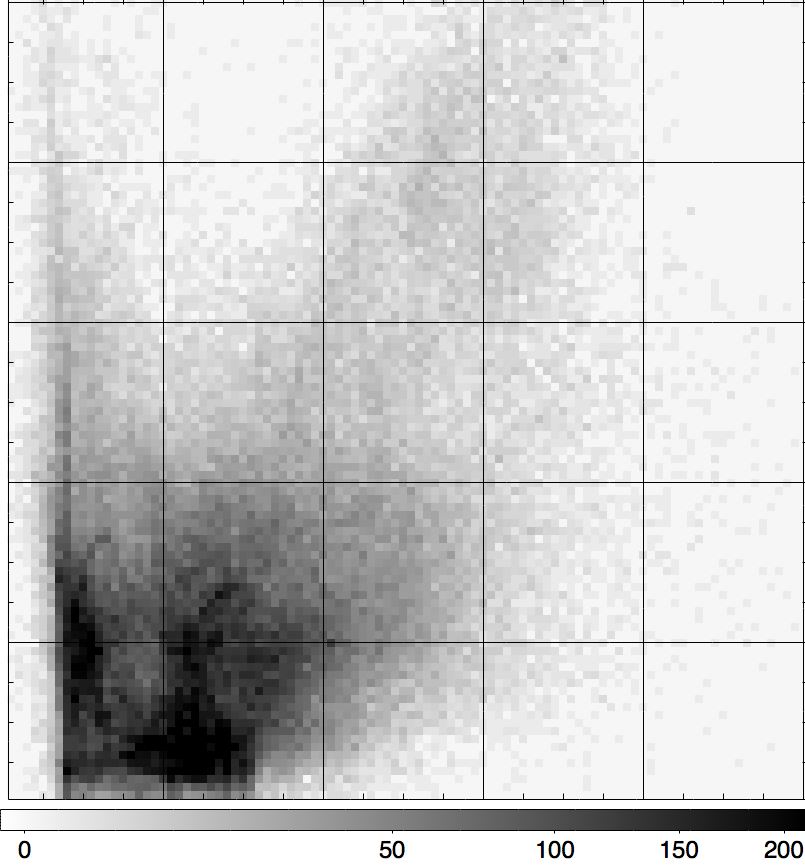}
\caption{Map of the distribution of the measured dust density $n_\mathrm{H}$ (vertical axis, linear scale from 0 to $5 \times 10^{55}\,\mathrm{H}\,\mathrm{arcsec}^{-2}$) as a function of the measured polarization degree $p_{\rm L}$ (horizontal axis, linear scale from 0 to 50\%). The color coding is proportional to the number of pixels in each $(p_{\rm L}, n_\mathrm{H})$ bin. \label{pl_over_density}}
\end{figure}

As discussed in Sect.~\ref{nebulathickness}, we based our reasoning on the hypothesis that the light scattering dust layer in the nebula of RS\,Pup is optically thin. A test of this hypothesis can be done by comparing the derived column density values $n_\mathrm{H}$ to the polarization degree $p_{\rm L}$ for each point in the nebula. For an optically thin layer and an overall random dust distribution, these two quantities should be independent of each other. Fig.~\ref{pl_over_density} shows the two-dimensional distribution of the number of pixels within each $(p_{\rm L}, n_\mathrm{H})$ bin. The small number of pixels with polarization values below $\approx 3$\% (horizontal axis) is caused by the masking of the central part of the nebula (where the polarization is the lowest) that presents artefacts due to the brightness of the central star. Overall, this diagram appears reasonably well populated, which validates our thin layer hypothesis. We note the presence of a population of locations on the nebula with simultaneous high polarization degree ($\approx 30\%$) and high column density ($\approx 4 \times 10^{55}\,\mathrm{H}\,\mathrm{arcsec}^{-2}$), in the upper right part of the diagram. They correspond essentially to the ridge area to the south of RS\,Pup, as well as to other dense areas to the northeast of the star.

%______________ Figure

\subsection{The southern ridge}

The maximum densities are reached over the ridge-like structure located to the south of RS\,Pup, at $n_\mathrm{H} = 10^{56}\,\mathrm{H}\,\mathrm{arcsec}^{-2}$.
%
%______________ Figure
\begin{figure}[]
\centering
% 700 pix, zoom 4, characters 24 Helvetica, center (1330, 1030), 174 pixels width
\includegraphics[width=4.4cm]{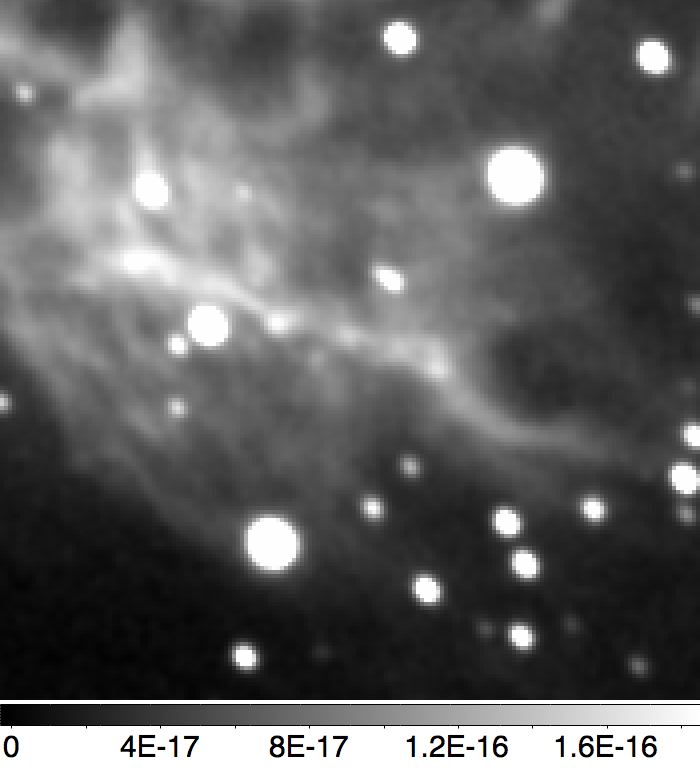}
\includegraphics[width=4.4cm]{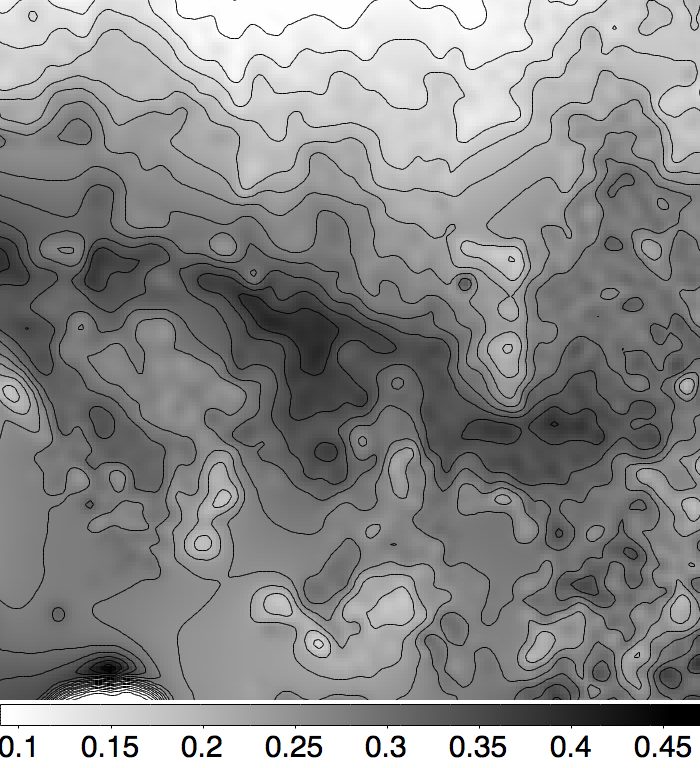}
\includegraphics[width=4.4cm]{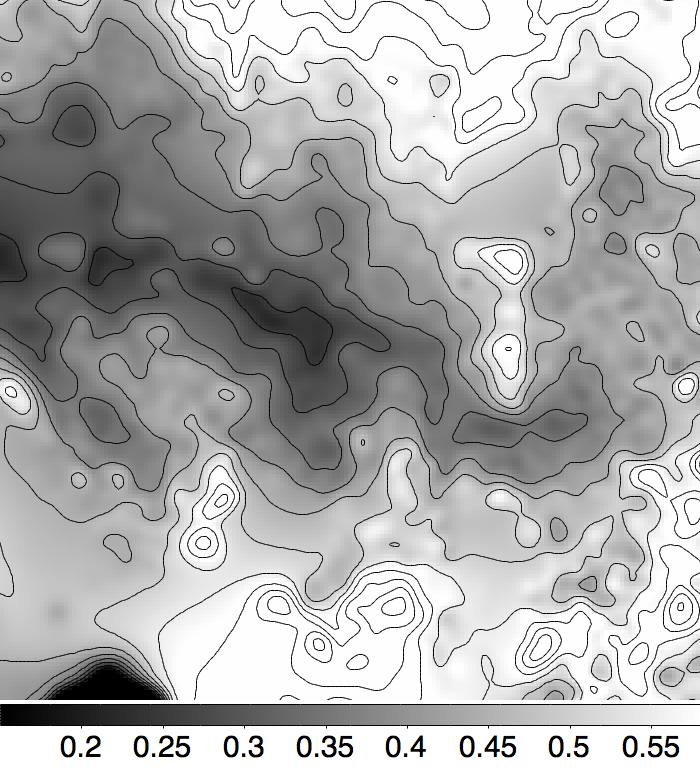}
\includegraphics[width=4.4cm]{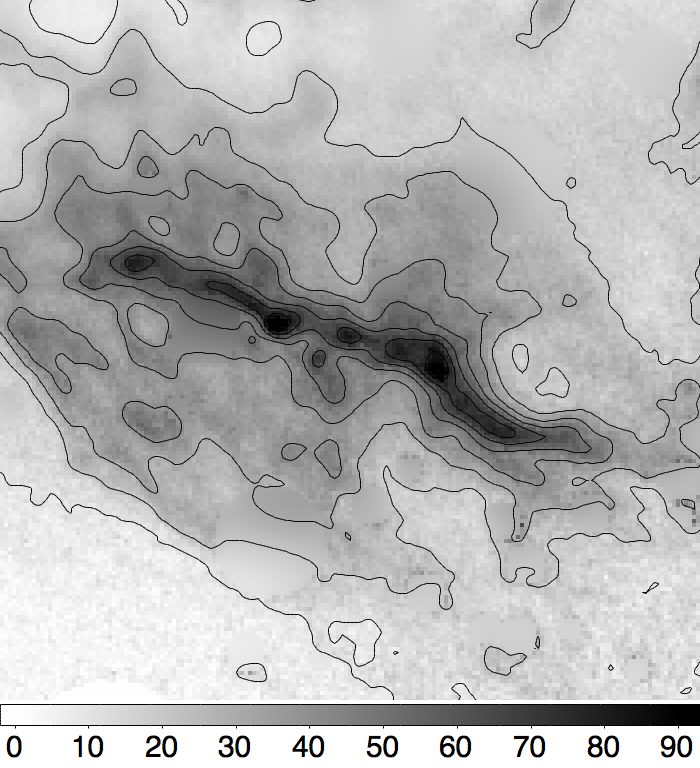}
\caption{Enlargement of the ridge nebular feature located south of RS\,Pup. {\it Top row:} EMMI intensity image (left, W\,m$^{-2}$\,$\mu$m$^{-1}$\,sr$^{-1}$, linear scale) and $p_{\rm L}$ map (right, with 2.5\% spacing contours starting at 0). {\it Bottom row:} altitude of the dust layer (left, in pc, with 0.05\,pc contours) and dust column density (right, $10^{54}\,\mathrm{H}\,\mathrm{arcsec}^{-2}$, with 10 unit contours). The field of view is $44 \times 44\arcsec$, corresponding to $0.4 \times 0.4$\,pc at the distance of RS\,Pup. \label{ridge}}
\end{figure}
An enlargement of this region of the nebula is presented in Fig.~\ref{ridge}. This roughly linear nebular feature presents relatively high $p_{\rm L}$ values, and is therefore located close to the plane of the sky at RS\,Pup's distance.
It is interesting to remark that this does not correspond to the photometrically brightest parts of the nebula, that reach only $n_\mathrm{H} \approx 5.0 \times 10^{54}\,\mathrm{H}\,\mathrm{arcsec}^{-2}$. There is also a slight spatial shift between the areas of maximum dust density and the lowest altitudes relative to the plane of the sky (Fig.~\ref{Zimage_pc}). The ridge structure therefore appears as a dense ribbon of dust, inclined with respect to the plane of the sky.

%__________________________________Discussion
\section{Discussion \label{discussion}}

\subsection{Total mass of the light-scattering material\label{totalmass}}

As described in Sect.~\ref{scatter}, we deduced the surface density of the dust from the scattered light intensity and the geometry of the scattering layer. This provides us with a means to estimate the total mass of the dust layer, by integrating over the nebula. We consider in our integration only the parts of the nebula that are visible in Fig.~\ref{nH_density}. We prefer not to extrapolate the density at the positions where we do not have a measurement, as the dust distribution is inhomogeneous, and this could result in a bias. We obtain a total number of H nucleons of $n_\mathrm{H} = 4.1 \times 10^{59} \pm 15\%$ within $1.8\arcmin$ from RS\,Pup (i.e. $\approx 200$\,kAU, or 1\,pc at the distance of RS\,Pup).

For their Milky Way dust model with $R_V = 3.1$, Weingartner \& Draine~(\cite{weingartner01}) find a total volume per H atom of $\tilde{V_\mathrm{g}}=2.26 \times 10^{-33}$\,m$^3$\,H$^{-1}$ for carbonaceous dust and $\tilde{V_\mathrm{s}}=3.94 \times 10^{-33}$\,m$^3$\,H$^{-1}$ for silicate dust (assuming a total carbon abundance per H nucleus $b_C = 6 \times 10^{-5}$, which is their favored value).
Considering the small relative abundance of carbonaceous dust compared to silicates, we neglect its contribution to the total dust mass, and we use the dust grain volume $\tilde{V_\mathrm{s}}$ in our computations.
Moreover, the contribution of the very small carbonaceous dust grains is essentially observable at infrared and microwave wavelengths, and is negligible in our visible scattered light observations.
Following Weingartner \& Draine~(\cite{weingartner01}), we adopt a density of 3500\,kg\,m$^{-3}$ for the silicate dust, which is an intermediate value between crystalline forsterite (3210\,kg\,m$^{-3}$) and fayalite (4390\,kg\,m$^{-3}$).

Based on our measurement of the total number of H nucleons, we obtain a total dust volume of $V_\mathrm{dust}=1.62 \times 10^{27}$\,m$^3$, and a total light-scattering dust mass of $M_\mathrm{dust}= (5.7 \pm 1.7) \times 10^{30}\,\mathrm{kg}=2.9 \pm 0.9\,M_\odot$. Taking into account the modeling uncertainties, we adopt a systematic uncertainty on this value of $\pm 30\%$.
We can now evaluate the total gas and dust mass of the circumstellar envelope of RS\,Pup by assuming the dust-to-gas ratio of $M_\mathrm{dust}/M_\mathrm{gas} = 1\%$ derived by Draine \& Li~(\cite{draineli07}) for the Milky Way. Draine et al.~(\cite{draine07}) compared this model to the dust observed in a sample of galaxies, and from their results we evaluate the uncertainty on this parameter to $\approx 30\%$. We therefore estimate the total mass of the nebula within $1.8\arcmin$ of RS\,Pup to $M_\mathrm{gas+dust} = 290\,M_\odot \pm 40\%$, i.e. approximately between 180 and $420\,M_\odot$.
As a remark, it is possible to compute directly $M_\mathrm{gas+dust} = n_\mathrm{H} \times m_\mathrm{H}$ with $m_\mathrm{H} = 1.672 \times 10^{-27}$\,kg, that gives a value of $345\,M_\odot$. With a dust to gas ratio of 1\%, we obtain a dust mass of $3.5\,M_\odot$, in agreement with our value of $M_\mathrm{dust}=2.9 \pm 0.9\,M_\odot$ derived above.

The total mass $M_\mathrm{gas+dust} = 290\,M_\odot \pm 40\%$ should be considered a lower limit for the true mass of the interstellar cloud in which RS\,Pup is embedded, for two reasons: 1) we do not include the faint extensions of the nebula that are located farther than $1.8\arcmin$ from the star (see Sect.~\ref{nebshape}), and 2) we are sensitive only to the dust that is located between us and RS\,Pup. The material located behind the plane of the sky at the distance of RS\,Pup remains invisible to us, due to the low efficiency of backwards scattering (Fig.~\ref{scatter_model}). Under the hypothesis that the dust layer has a symmetric counterpart behind RS\,Pup, the true total mass of the cloud could be the double of the above figures, and significantly more if we include the faint extensions located beyond $1.8\arcmin$.

An estimate of the dust mass around RS\,Pup was computed by Havlen~(\cite{havlen72}). He obtains a dust mass of $M_\mathrm{dust} = 0.4\,M_\odot$ based on a simple multiple shell model, and a second value $M_\mathrm{dust} = 0.03\,M_\odot$ using a more indirect approach (as outlined by Lynds~\cite{lynds68}). These estimates are both lower than ours, but his first value is relatively comparable. As shown in Fig.~\ref{westerlund_emmi}, many dust features are difficult to observe at relatively large separations on the original photographic plates, and this probably caused an underestimation of the total mass by Havlen. Moreover, faint parts of the nebula, not counted in Havlen's estimate, exhibit a particularly high dust density (e.g. the ridge located south of RS\,Pup).

\subsection{Shape of the nebula \label{nebshape}}

\begin{figure}[ht]
\centering
\includegraphics[width=\hsize]{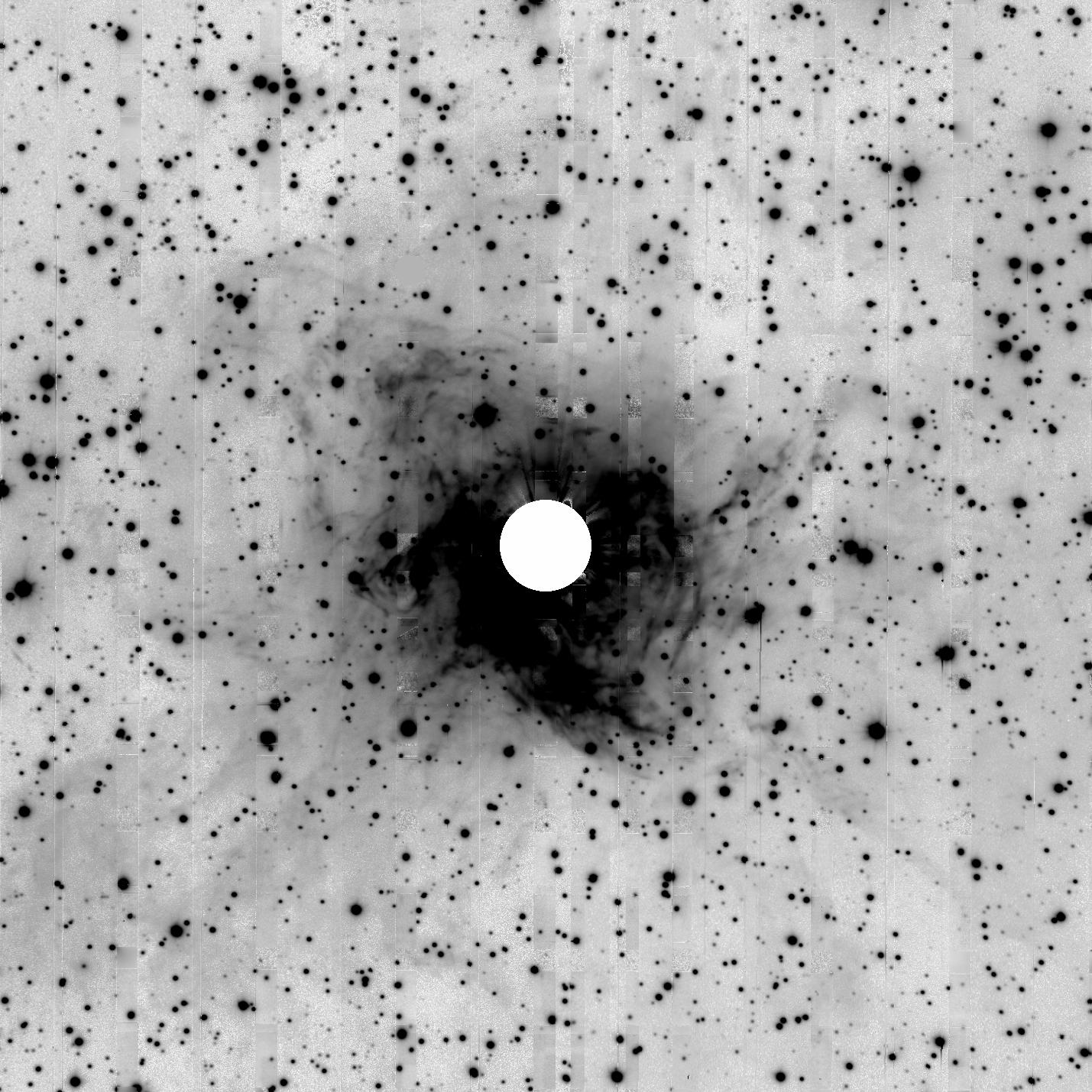}
\caption{Combined FORS $V$ band image of RS\,Pup. The field of view is $6\arcmin \times 6\arcmin$, with North up and East to the left.\label{combinedFORS}}
\end{figure}

The light-scattering material surrounding RS\,Pup appears to be spread over an irregular surface (Fig.~\ref{Zimage_pc}). This geometry does not present a well defined central symmetry relative to the Cepheid. Its density is inhomogeneous, with high dust densities over some sections of the nebula (e.g. along the ridge located south of RS\,Pup), and mostly void areas (e.g. in the northwestern part).
We present in Fig.~\ref{combinedFORS} the result of the direct co-addition of our FORS images (\#1 to 8 and 12 to 19 in Table~\ref{fors_log}), each exposed $4 \times 90$\,s, giving a total exposure time of 1.6\,h. Faint extensions are detectable up to at least $3\arcmin$ from the star, and it is not excluded that the nebula extends even further away. The background appears slightly non uniform due to the intensity scale that is chosen to enhance the faintest extents of the nebula. The overall geometry of the nebula appears to have two preferential axes, towards the northeast-southwest and northwest-southeast of the central star.

%______________ Figure

As proposed in Paper~I, the nebula has likely been shaped by the past evolution of RS\,Pup in a pre-existing, higher density interstellar cloud.
About $\approx$10 to 20\,Myr in the past, RS\,Pup went through a phase during which it was a massive, most probably fast rotating B type dwarf. During this period, the strong equatorial and polar winds of the star combined with its energetic photons to sweep the interstellar dust away from the star. This effect may have compressed the dust into the thin veil that we observe today. The current very high intrinsic brightness of the Cepheid probably also contributes to the shaping of the nebula, but at a slower rate. The closest dust to RS\,Pup that we observe in our FORS images is located at linear distances of $\approx 1$ to 2\,pc from the star, which is already relatively far away. The fact that we do not observe a physical displacement of the nebular features over several decades (Appendix~\ref{tempevol}) is consistent with this large separation between the star and the dust features.
Such cavities carved by winds and/or radiation pressure have been studied for massive hot stars and supernova remnants (see e.g. Lozinskaia~\cite{lozinskaya92}), but not so much for regular giant stars. It is interesting to remark that the creation of cavities by stellar winds and radiation create hydrodynamic instabilities and turbulence that seem to be present on the surface of RS\,Pup's nebula as wisps and curls.

Another important phase of RS\,Pup's evolution was when it was a red supergiant (RSG), after its first crossing of the instability strip and before entering again the instability strip as a Cepheid. A plausible scenario could therefore be that mass-loss took place during the RSG phase of RS\,Pup, leading to the formation of dust relatively close to the star. A good match of the position of RS\,Pup in the HR diagram a few 100\,000\,years in the past is provided by Betelgeuse, as the mass of this RSG, although uncertain, is comparable to that of RS\,Pup: Neilson \& Lester~(\cite{neilson11}) recently proposed a mass of $11.6^{+5.0}_{-3.9}\,M_\odot$ for Betelgeuse (but Dolan et al.~\cite{dolan08} obtained $\approx 21\,M_\odot$), to be compared with $12.8 \pm 0.9\,M_\odot$ for RS\,Pup (Caputo et al.~\cite{caputo05}). The subsequent re-heating of the stellar surface after the RSG phase would have resulted in the disappearance of the dust close to the star, and the sweeping of the interstellar medium.
Considering the thinness (Sect.~\ref{nebulathickness}) of the observed dust layer, the hypothesis of a sweeping of the interstellar dust during the main sequence and/or RSG phases of RS\,Pup appears compatible with the observations.

\subsection{Origin of the nebula}

%In this Section, we briefly examine several possible origins for the dust present around RS\,Pup: interstellar material predating the star, mass-loss from the Cepheid itself (either in the past or presently), and mass-loss from an unseen companion. We leave a more detailed study of the possible binarity and other physical properties of RS\,Pup itself for a forthcoming paper.

The high mass derived for the nebula is clearly incompatible with the scenario that RS\,Pup created its circumstellar nebula through mass loss. The bulk of the material observed in scattered light therefore appears to be of interstellar origin. The higher density environment in which RS\,Pup is presently located could either be the remnant of the molecular cloud from which the star formed, or unrelated interstellar material into which RS\,Pup would be temporarily embedded, due to its proper motion in the Galaxy. On the latter scenario, it is interesting to notice that Westerlund~(\cite{westerlund63}) pointed out that the reflection nebula surrounding RS\,Pup is a part of a stellar association consisting of early-type stars. Although the physical presence of RS\,Pup in this association is possible, the very young age of the association ($\approx 4$\,Myr) compared to the Cepheid may indicate that the Cepheid did not form within this association. A more detailed discussion on this question can be found in Paper~I.

However, a fraction of the nebular material could still originate in past or present stellar mass-loss from RS\,Pup. As discussed previously, mass-loss probably happened during the RSG phase that took place in the recent evolution of the star. In addition, Kervella et al.~(\cite{kervella09}) identified a compact and hot circumstellar envelope around RS\,Pup, that is likely due to presently ongoing mass loss. This also supports the findings of Deasy~(\cite{deasy88}), who identified a very significant mass loss rate for RS\,Pup ($\dot{M} = 10^{-6}\,M_{\odot}\,{\rm yr}^{-1}$), based on IUE spectra. 

The strong changes in the pulsation period of RS\,Pup (Berdnikov et~al.~\cite{berdnikov09}) might also indicate presently ongoing mass loss. The $O-C$ diagram of RS\,Pup (Fig.~1 in Berdnikov et~al.~\cite{berdnikov09}) indicates sudden and strong fluctuations superimposed on the secular increase in the pulsation period. The secular variations indicate period lengthening at a rate of 137 s/year, while the most recently superimposed fluctuation in the pulsation period corresponds to a sudden change by 0.15\% at about JD\,2\,452\,000. The recent adaptive optics observations by Gallenne et al.~(\cite{gallenne11}) also indicate that the close environment of RS\,Pup contains a significant quantity of hydrogen. The observed contribution of this envelope in the near-infrared hydrogen lines is significant, and points at a circumstellar location.

There is another piece of evidence that might refer to the origin of some of the material surrounding RS\,Pup. The radial velocity observations available in the literature indicate variations in the center-of-mass radial velocity of RS\,Pup as if this Cepheid belonged to a spectroscopic binary system. The companion could trigger an enhancement of the mass loss from the low gravity pulsating atmosphere of the supergiant Cepheid variable. This companion must have a significantly lower mass than RS\,Pup, as the flux contribution from a more massive star would either be easily detectable, or it would have already exploded as a supernova (and the nebula of RS\,Pup does not present the characteristics of a supernova remnant).

%__________________________________Conclusion
\section{Conclusion}

From a combination of polarimetric and direct imaging of RS\,Pup's nebula, we derived the geometry of the light-scattering dust layer surrounding the star, as well as a map of its column density. The observed distribution of the dust appears irregular and not centered on the Cepheid. The dust mass of the nebula is estimated to $M_\mathrm{dust} = 2.9 \pm 0.9\,M_\odot$, and its total mass to $M_\mathrm{gas+dust} = 290\,\pm 120\,M_\odot$. These high values exclude the possibility that the bulk of the material of the nebula was ejected by the Cepheid itself. The lack of central symmetry in the nebula also points at an interstellar origin for the nebular material. Following Kervella et al.~(\cite{kervella09}), we propose that the nebula is either a remnant of the interstellar cloud in which RS\,Pup formed, or a particularly dense interstellar cloud in which it is temporarily embedded due to its proper motion.
There are indications in the nebula geometry that the Cepheid participated in the shaping of the dust cloud, e.g. through its radiative pressure or wind. The Cepheid may also have contributed a fraction of the material present in the nebula, in particular during a past RSG phase, but this fraction must be small compared to the interstellar material, considering the involved dust and gas masses.

The presence of the large dusty circumstellar nebula around RS\,Pup therefore does not appear to be in itself a property of the star, in the sense that it is not the result of stellar mass-loss by the Cepheid. Instead, the position of RS\,Pup in the nebula appears essentially coincidental, thus providing a natural explanation to the scarcity of similar Cepheid-nebula associations. Although bow shocks have been observed by Marengo et al.~(\cite{marengo10}) in the $\delta$\,Cep system, and infrared excess has been identified by Barmby et al.~(\cite{barmby11}) in several Cepheids, the nature of the envelope of RS\,Pup appears fundamentally different, both in extension and total mass.
Another consequence is that the dusty nebula is not a good candidate to resolve a possible evolutionary vs. pulsational mass discrepancy for RS\,Pup (see e.g. Keller~\cite{keller08}, Caputo et al.~\cite{caputo05}, and Neilson \& Lester~\cite{neilson08}, \cite{neilson09}). However, the compact and hot envelope observed by Kervella et al.~(\cite{kervella09}) still appears as a plausible contributor.

Our next step, that will be the subject of a forthcoming article, is to determine the distance of RS\,Pup using its light echoes, as the geometry of the light-scattering dust is now well determined.

%__________________________________Acknowledgements
\begin{acknowledgements}
We thank Dr. D. L. Welch for providing the CFHT images of RS\,Pup presented in Appendix~\ref{tempevol}.
This research received the support of PHASE, the high angular resolution
partnership between ONERA, Observatoire de Paris, CNRS and University Denis Diderot Paris 7.
We acknowledge financial support from the ``Programme National de Physique Stellaire" (PNPS) of CNRS/INSU, France
and from the C98090 PECS project of the European Space Agency.
We also took advantage of the SIMBAD and VIZIER databases at the CDS, Strasbourg (France),
NASA's Astrophysics Data System Bibliographic Services, and the facilities of the Canadian Astronomy Data Centre
(operated by the National Research Council of Canada with the support of the Canadian Space Agency).
In memoriam Dr. B. E. Westerlund (1921-2008), discoverer of the nebula of RS Pup.
\end{acknowledgements}

%__________________________________Bibliography
{}

\begin{appendix}
\section{Search for temporal evolution of RS\,Pup's nebular features \label{tempevol}}

\subsection{Westerlund's photographic plates 1960-1963}

Soon after his discovery of the nebula surrounding RS\,Pup, Westerlund obtained a series of photographic plates with the 74-inch Mount Stromlo telescope (located near Canberra, Australia) to search for the propagation of light echoes (Westerlund~\cite{westerlund61}). These plates were subsequently used by Havlen for his pioneer study of the nebula (Havlen~\cite{havlen72}), and a detailed description of the plates can be found in this publication. For the present work, we digitized the original photographic plates, that were preserved by one of us (RJH) for almost 50\,years, using a flatbed scanner at a resolution of 2400\,dpi. This optical resolution is sufficient to resolve the film grain. We matched the astrometric WCS of the scanned plates with the EMMI images obtained in 2007 using a dozen of background stars spread over the image field. The proper motion of a few bright stars of the field is noticeable, however, the matching of the images is satisfactory.

We carefully blinked the Westerlund plates with the EMMI observations presented in Paper~I. The EMMI images were interpolated at the phase of the Westerlund observations using the procedure described in Sect.~\ref{emmiobs}. Among the 18 epochs covering the 1960-1963 period, we focused our comparison on three epochs that show particularly well the circumstellar nebula: 30 March 1962 (visual), 5 May 1962 (visual) and 26 February 1963 (blue and visual). These plates were obtained under very good seeing conditions, and are the best of the series. We did not detect any change in the morphology of the nebula. This negative result was expected considering the fact that a $1\arcsec$ angular displacement corresponds to a linear displacement of $\approx 2000$\,AU at the distance of RS\,Pup. Over 50\,years, such a $1\arcsec$ displacement would therefore require a very fast linear velocity of at least $\approx 200$\,km/s.

%______________ Figure
\begin{figure*}[ht]
\centering
\includegraphics[width=\hsize]{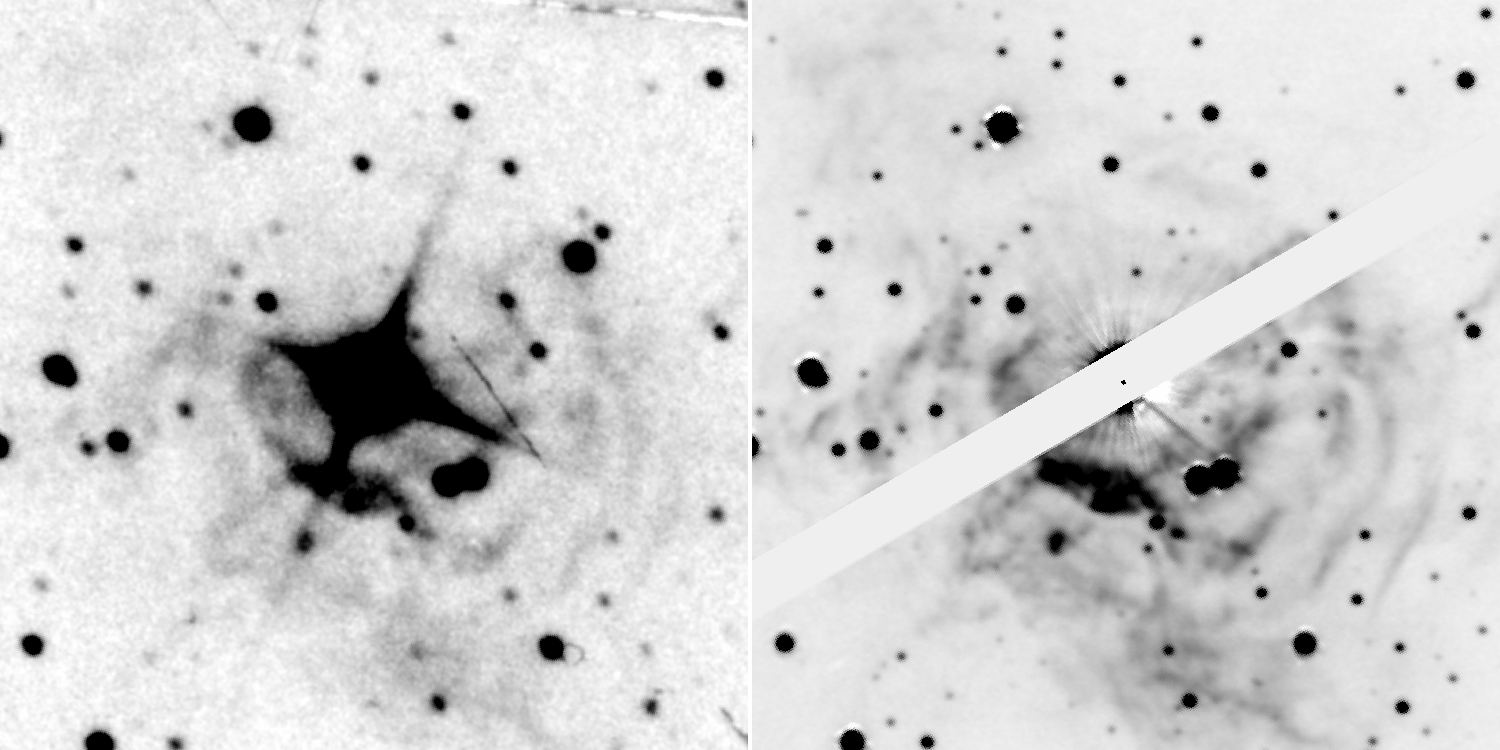}
\caption{Comparison of Westerlund's photographic plate \#3117 with the interpolated and PSF-subtracted EMMI observation at the same phase ($\phi=0.84$). The field of view is $1.4\arcmin \times 1.4\arcmin$, with North up and East to the left. \label{westerlund_emmi}}
\end{figure*}
%
%______________ Figure
\begin{figure*}[ht]
\centering
\includegraphics[width=9cm]{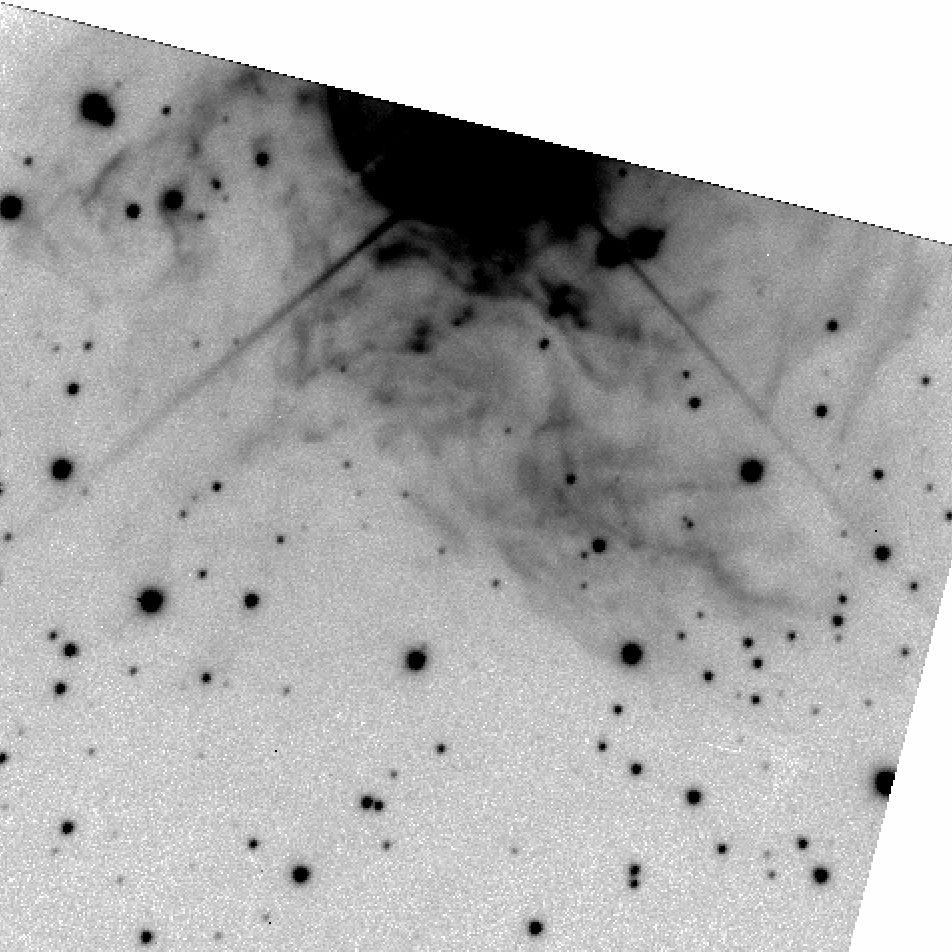}
\includegraphics[width=9cm]{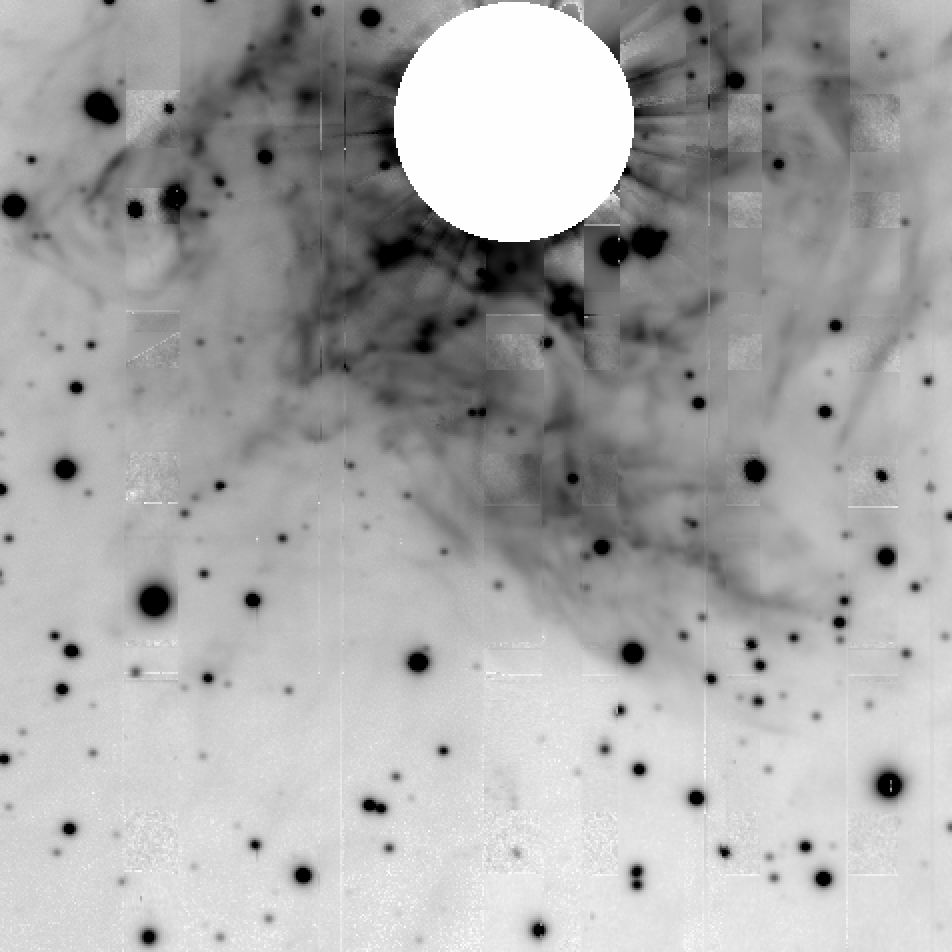}
\caption{CFHT/HRC image of the circumstellar nebula of RS Pup (left, 1992) and the same field observed with FORS (right, 2009). The field of view is $2\arcmin \times 2\arcmin$, with North up and East to the left.\label{rspup_cfht}}
\end{figure*}

\subsection{CFHT observations 1992-1993}

The circumstellar nebula of RS\,Pup was observed on the nights of 4-6 January 1992, 9 April 1992, and 28 March 1993, using the HRC instrument installed at the prime focus of the Canada-France-Hawaii Telescope, for Michael J. Pierce \& Douglas L. Welch. Exposures of 30 to 600\,s were obtained in the $B$, $V$ and $R$ bands. We present in Fig.~\ref{rspup_cfht} the best image in the $V$ band (06-01-1992), compared to our second epoch observation with FORS (27-05-2009). The two observations do not correspond to the same phase, but we could not detect any change in the morphology of the nebula (apart from the photometric variations due to the echoes).
%
%______________ Figure
\begin{figure}[ht]
\centering
\includegraphics[width=4.4cm]{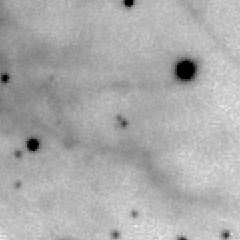}
\includegraphics[width=4.4cm]{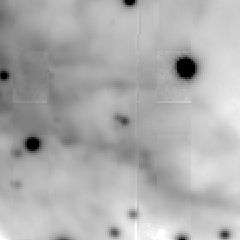}
\caption{Displacement of a faint star in the field surrounding RS\,Pup between 1992 (left) and 2009 (right). The displacement of the brighter of the two stars near the center of the field is of $\approx0.6\arcsec$. The field of view is $30\arcsec \times 30\arcsec$.\label{starmotion}}
\end{figure}
However, we observe the apparent displacement of several stars in the field over the 17\,years separating the two observations (an example is presented in Fig.~\ref{starmotion}). The bright pair of point-like sources visible above the center of the FORS image (close to the central vertical white stripe) is an artefact of the instrument.

\end{appendix}

\end{document}